\documentclass[12pt]{article}
\usepackage{epsfig}
\usepackage{color}
\usepackage{amssymb,amsmath}
\begin{document}

\thispagestyle{empty}

{\hspace*{-0.6cm}DESY 10-023} \\
{February 2010} \\

\vspace*{1cm}

\begin{center}
\begin{Large}
\begin{bf}
Errors in Measuring Transverse and Energy Jitter \\
\vspace{0.3cm}
by Beam Position Monitors
\end{bf}
\end{Large}

\vspace{1cm}

\begin{large}
V. Balandin, W. Decking and N. Golubeva
\end{large}

\vspace{0.3cm}

\begin{small}
\begin{it}
Deutsches Elektronen-Synchrotron DESY, \\
Notkestrasse 85, 22607 Hamburg, Germany
\end{it}
\end{small}

\end{center}

\vspace{0.8cm}

\begin{abstract}
The problem of errors, arising due to finite 
BPM resolution, in the difference orbit parameters, which are found as 
a least squares fit to the BPM data, is one of the standard and important
problems of accelerator physics.
Even so for the case of transversely uncoupled motion the covariance
matrix of reconstruction errors can be calculated ``by hand",
the direct usage of obtained solution, as a tool for designing of 
a ``good measurement system", does not look to be fairly straightforward. 
It seems that a better understanding of the nature of the problem is 
still desirable. We make a step in this direction introducing dynamic
into this problem, which at the first glance seems to be static. 
We consider a virtual beam consisting of virtual particles obtained 
as a result of application of reconstruction procedure to ``all possible values'' 
of BPM reading errors. This beam propagates along the beam line according 
to the same rules as any real beam and has all beam dynamical characteristics,
such as emittances, energy spread, dispersions, betatron functions and etc.
All these values become the properties of the BPM measurement system. 
One can compare two BPM systems comparing their error emittances and rms 
error energy spreads, or, for a given measurement system, one can achieve 
needed balance between coordinate and momentum reconstruction errors by matching 
the error betatron functions in the point of interest to the desired values.
\end{abstract}

\clearpage
\setcounter{page}{2}
\tableofcontents

\clearpage
\section{Introduction}

\hspace{0.5cm}
The determination of variations in the transverse 
beam position and in the beam energy using readings of 
beam position monitors (BPMs) is one of the standard and important problems
of accelerator physics. If the optical model of the beam line and BPM 
resolutions are known, the typical choice is to let jitter parameters 
be a solution of the weighted linear least squares problem. 
Even so for the case of transversely uncoupled motion this least squares 
problem can be solved ``by hand", the direct usage of obtained analytical 
solution as a tool for designing of a ``good measurement system" does not 
look to be fairly straightforward. It seems that a better understanding 
of the nature of the problem is still desirable.

A step in this direction was made in the paper \cite{ThePaper}, where
dynamic was introduced into this problem which in the beginning 
seemed to be static. When one changes the position of the reconstruction 
point, the estimate of the jitter parameters propagates along the beam
line exactly as a particle trajectory and it becomes possible (for every fixed
jitter values) to consider a virtual beam consisting from virtual particles 
obtained as a result of application of least squares reconstruction procedure 
to ``all possible values'' of BPM reading errors. The dynamics of the centroid 
of this beam coincides with the dynamics of the true difference orbit 
and the covariance matrix of the jitter reconstruction errors can be treated
as the matrix of the second central moments of this virtual beam
distribution.

In accelerator physics a beam is
characterized by its emittances, energy spread, dispersions, betatron
functions and etc. All these values immediately become the properties
of our BPM measurement system. From now one can compare two BPM systems
comparing their error emittances and error energy spreads, or, for a
given measurement system, one can achieve needed balance between coordinate
and momentum reconstruction errors by matching the error betatron functions 
in the point of interest to the desired values.

This dynamical point of view on the BPM measurement system was explored
in \cite{ThePaper} in application to the case of transversely uncoupled
nondispersive beam motion and in this paper we continue this study
adding energy degree of freedom.\footnote{It is clear, that such
considerations, if needed, can also be done for the case of the fully coupled 
six dimensional motion. It is also clear that in similar fashion one can 
approach some other problems connected with the error propagation. 
It should not be necessary the BPM reading errors, it could be, for example,   
errors in the kick angles produced by the orbit feedback system.}
The paper by itself is organized as follows. 
In section 2 we introduce all needed notations,
formulate the problem and give its
standard least squares solution. As a new element,
we formulate the necessary and sufficient conditions  
for the BPM system to be able to distinguish between
transverse and energy jitters
in terms of its three BPM subsystems.
In section 3 (the core section of this paper) we make
parametrization of the covariance matrix of the jitter
reconstruction errors using the usual accelerator physics
concepts of emittance, energy spread, dispersion and
betatron functions. We also show that the error dispersion is not 
simply one of the many dispersions which could propagate through our beam
line. It, in analogy with the error betatron functions \cite{ThePaper},
is by itself solution of some minimization problem and is uniquely
determined by transport matrices between BPM locations and by
BPM resolutions. In section 4 we consider the measurement system
which utilizes three beam position monitors (the minimum number
of BPMs needed) and analyze in details effect of 
symmetries of the optics between BPM locations.
In section 5 we continue the investigation
of periodic measurement systems started in \cite{ThePaper}. 
This time with the main accent on
achievable energy resolution. And, finally, in section 6 we discuss
application of the Courant-Snyder quadratic form as error estimator,
even so in the case when energy degree of freedom is taken into
account this quadratic form is not bound to be an invariant.

\section{Problem and Its Least Squares Solution}

\hspace{0.5cm}
Let us consider a magnetostatic beam line which is built from optical elements
which are symmetric about the horizontal midplane $\,y = 0$.
In such magnetic system the transverse particle motion is uncoupled
in linear approximation, the vertical oscillations are dispersion free and
errors in reconstruction of their parameters
were already studied in \cite{ThePaper}, and in this paper we will examine 
together $x$-plane and energy degrees of freedom because they are connected
through (linear) dispersion. 

We will use the variables 
$\;\vec{z} = (x, \, p, \, \varepsilon)^{\top}\;$ 
for the description of the horizontal dispersive beam motion.
Here, as usual, $\,x\,$ is the horizontal particle coordinate, $\,p\,$ is the
horizontal canonical momentum scaled with the kinetic momentum of the
reference particle and the variable $\,\varepsilon\,$ stays for
the relative energy (or momentum) deviation.\footnote{The exact form of the variable 
$\,\varepsilon\,$ which we have in mind can be found in \cite{HamMethod}, 
but let us note that for the present study the particular form of this
variable is unimportant. 
Let us also note that while in \cite{ThePaper} the symbol
$\,\varepsilon\,$ was used for the BPM reading errors, in this paper we prefer
to use it for the  relative energy deviation, 
and for the BPM reading errors we will introduce $\,\varsigma\,$ as new notation.} 
As orbit parameters we will understand  values of 
$\,x,\, p\,$ and $\,\varepsilon\,$ given in some predefined point 
in the beam line (reconstruction point with longitudinal position $\,s = r\,$) 
and as transverse and energy jitter in this point
we will mean the difference 

\noindent
\begin{eqnarray}
\delta \vec{z}(r) \,=\,
\left(\,\delta x(r),\, 
\delta p(r),\, 
\delta \varepsilon(r)\,\right)^{\top} 
=\,
\left(\,
x(r) \,-\, \bar{x}(r),\;
p(r) \,-\, \bar{p}(r),\;
\varepsilon \,-\, \bar{\varepsilon}\,
\right)^{\top}
\label{INT_1}
\end{eqnarray}

\noindent
between parameters of the instantaneous orbit
and parameters of some predetermined reference (golden) trajectory 
$\;(\bar{x},\, \bar{p},\, \bar{\varepsilon})^{\top}$.

Let us assume that we have $n$ BPMs in our beam line placed at positions 
$\,s_1, \ldots, s_n\,$ and they deliver readings

\noindent
\begin{eqnarray}
\vec{b}_{c} \;=\; \left(\,b_1^c, \, \ldots \, b_n^c \, \right)^{\top}
\label{SEC1_1}
\end{eqnarray}

\noindent
for the current trajectory with previously recorded 
observations for the golden orbit being 

\noindent
\begin{eqnarray}
\vec{b}_{g} \;=\; \left(\,b_1^g, \, \ldots \, b_n^g \, \right)^{\top}\,.
\label{SEC1_2}
\end{eqnarray}

\noindent
Suppose that the difference between these readings 
can be represented in the form

\noindent
\begin{eqnarray}
\delta \vec{b}_{\varsigma}
\; \stackrel{\rm def}{=} \;
\vec{b}_{c} \,-\, \vec{b}_{g} \; = \;
\left(
\begin{array}{c}
x(s_1) - \bar{x}(s_1) \\
\vdots                \\
x(s_n) - \bar{x}(s_n)
\end{array}
\right) \;+\; \vec{\,\varsigma} \,,
\label{SEC1_3}
\end{eqnarray}

\noindent
where the random vector 
$ \, \vec{\,\varsigma} \, = \, ( \,\varsigma_1, \, \ldots, \,\varsigma_n\,)^{\top}\,$ 
has zero mean and positive definite covariance matrix $\,V_{\varsigma}\,$, i.e. that

\noindent
\begin{eqnarray}
\big< \, \vec{\,\varsigma} \, \big> \; = \; \vec{\,0} , 
\hspace{0.7cm}
{\cal V} \left(\, \vec{\,\varsigma} \, \right) \;=\;
\big< \, \vec{\,\varsigma} \cdot \vec{\,\varsigma}^{\top} \big>
\,-\, 
\big< \, \vec{\,\varsigma} \, \big> \cdot
\big< \, \vec{\,\varsigma} \, \big>^{\top}
 \;=\;V_{\varsigma} \;>\;0. 
\label{SEC1_4}
\end{eqnarray}

The purpose of this paper is to study the influence of 
BPM reading errors  $\,\vec{\,\varsigma}\,$ 
on precision of reconstruction of jitter parameters
under assumption that optical model of the beam line 
is known. The additional assumptions which we will make
are: the covariance matrix $\,V_{\varsigma}\,$ stays constant
and the BPM reading errors can be treated as independent  
from one measurement to the other.
So BPM errors that are correlated from measurement to measurement
(calibration and other systematic errors, drifting BPM readings and etc.) 
and fluctuations in BPM resolutions will be not considered.
In practical applications these assumptions may or may not be realistic,
but, first, they make the underlying mathematics almost 
trivial\footnote{Under these assumptions errors in the reconstruction 
process can be modeled as a sequence of independent identically distributed
random variables (like in coin tossing) and therefore all
probabilistic characteristics can be obtained studying errors in 
reconstruction of the result of only one
measurement, but for all possible values of $\,\vec{\,\varsigma}$.}
and, second, their satisfaction is, in some sense, one of the
goals for the BPM and BPM electronics designers. 

Let $\,A_m(r)\,$ be a transfer matrix from location of the
reconstruction point to the $m$-th BPM location

\noindent
\begin{eqnarray}
A_m(r) \, = \,
\left(
\begin{array}{lll}
a_m(r) & c_m(r) & g_m(r)\\
e_m(r) & d_m(r) & f_m(r)\\
0 & 0 & 1
\end{array}
\right),
\hspace{0.5cm}
a_m(r) \, d_m(r) - c_m(r) \, e_m(r) \, \equiv \, 1,
\label{m1}
\end{eqnarray}

\noindent
and let us assume that the Cholesky factorization 
$\,V_{\varsigma} = R_{\varsigma}^{\top} R_{\varsigma}\,$ 
of the covariance matrix $\,V_{\varsigma}\,$ is known.
As usual, we will find an estimate

\noindent
\begin{eqnarray}
\delta \vec{z}_{\varsigma} (r) \;=\;  
(\delta x_{\varsigma}(r),\, 
\delta p_{\varsigma}(r),\, 
\delta \varepsilon_{\varsigma}(r))^{\top} 
\label{m21}
\end{eqnarray}

\noindent
for the difference orbit parameters (\ref{INT_1}) in the presence of BPM reading errors
by solving the following weighted linear least squares problem

\noindent
\begin{eqnarray}
\min_{ \delta \vec{z}_{\varsigma} }  \;\;
\left\| M_{\varsigma} \cdot \delta \vec{z}_{\varsigma} \;-\; 
R_{\varsigma}^{-\top} \cdot \delta \vec{b}_{\varsigma} \,\right\|_2^2.
\label{SEC1_9}
\end{eqnarray}

\noindent
Here $\, \| \cdot \|_2 \,$ denotes the Euclidean vector norm,
$\,M_{\varsigma} \,=\, R_{\varsigma}^{-\top} M\,$ and

\noindent
\begin{eqnarray}
M \;=\; 
\left(
\begin{array}{ccc}
a_1(r) & c_1(r) & g_1(r)\\
\vdots & \vdots & \vdots\\
a_n(r) & c_n(r) & g_n(r)
\end{array}
\right).
\label{a2}
\end{eqnarray}

The problem (\ref{SEC1_9}) always has at least one solution 
and, if we will assume that the matrix $\,M_{\varsigma}\,$ has
full column rank $(\mbox{rank}(M_{\varsigma}) = 3)$, 
then the solution of this problem is unique 
and is given by the well known formula

\noindent
\begin{eqnarray}
\delta \vec{z}_{\varsigma}(r) \; = \;
\left(M_{\varsigma}^{\top}(r) M_{\varsigma}(r) \right)^{-1}
M_{\varsigma}^{\top}(r) \, R_{\varsigma}^{-\top} \cdot \delta \vec{\,b}_{\varsigma},
\;=
\nonumber
\end{eqnarray}

\noindent
\begin{eqnarray}
=\; \left(M^{\top}(r) V_{\varsigma}^{-1} M(r) \right)^{-1}
M^{\top}(r) \, V_{\varsigma}^{-1} \cdot \delta \vec{\,b}_{\varsigma}\,.
\label{SEC1_15}
\end{eqnarray}

\noindent
The calculation of the covariance matrix of the errors of this estimate 
(object of our main interest) is also standard and gives the following result

\noindent
\begin{eqnarray}
V_z(r) \; \stackrel{\rm def}{=} \;
{\cal V} \left(\, \delta \vec{z}_{\varsigma}(r) \, \right)
\;=\; 
\left(M_{\varsigma}^{\top}(r) M_{\varsigma}(r) \right)^{-1}
\;=\;
\left(M^{\top}(r) \, V_{\varsigma}^{-1} \, M(r) \right)^{-1}\,.
\label{SEC1_16}
\end{eqnarray}

Let us discuss in more details the important condition for the 
matrix  $\,M_{\varsigma}\,$ to have full column rank. 
This condition will allow us to separate betatron and dispersion oscillations
at the BPM locations and, therefore, will make our system applicable for measuring 
transverse and energy jitter.

Because the matrix $\,R_{\varsigma}\,$ is nondegenerated, the rank of the
matrix  $\,M_{\varsigma}\,$ is always equal to the rank of the matrix $\,M$,
and the matrix $\,M$, in the next turn, will have full column rank
if and only if the Gram determinant 
$\,\Gamma\left(\vec{a},\,\vec{c},\,\vec{g}\right)\,$ 
of its column vectors 

\noindent
\begin{eqnarray}
\vec{a} \;=\; \left(a_1, \ldots , a_n\right)^{\top}, \hspace{0.35cm}
\vec{c} \;=\; \left(c_1, \ldots , c_n\right)^{\top}, \hspace{0.35cm}
\vec{g} \;=\; \left(g_1, \ldots , g_n\right)^{\top}
\label{m2}
\end{eqnarray}

\noindent
is not equal to zero. 

To find desired expression for the Gram determinant
let us introduce $\,B_{mk}\,$ - transport matrix from the location of 
the BPM with index $m$
to the location of the BPM with index $k$

\noindent
\begin{eqnarray}
B_{mk} \; = \; A_k \, A_m^{-1} \; = \;
\left( 
\begin{array}{ccc}
\mbox{\ae}_{11}^{mk} & \mbox{\ae}_{12}^{mk} & \mbox{\ae}_{16}^{mk} \\
\mbox{\ae}_{21}^{mk} & \mbox{\ae}_{22}^{mk} & \mbox{\ae}_{26}^{mk} \\
0 & 0 & 1
\end{array}
\right).
\label{binC_1}
\end{eqnarray}

With these notations and using Binet-Cauchy formula one can obtain
after some straightforward manipulations

\noindent
\begin{eqnarray}
\Gamma\left(\vec{a},\,\vec{c},\,\vec{g}\right) \;=\;
\det \left(M^{\top} M\right) \;=\;
\sum\limits_{1 \leq i < j < k \leq n}
\left(\mbox{\ae}_{52}^{ij} \,\mbox{\ae}_{12}^{jk} \;-\; 
\mbox{\ae}_{12}^{ij} \,\mbox{\ae}_{16}^{jk}\right)^2 \;=
\nonumber
\end{eqnarray}

\noindent
\begin{eqnarray}
=\;
\frac{1}{6}
\sum\limits_{i,\,j,\,k = 1}^{n}
\left(\mbox{\ae}_{52}^{ij} \,\mbox{\ae}_{12}^{jk} \;-\; 
\mbox{\ae}_{12}^{ij} \,\mbox{\ae}_{16}^{jk}\right)^2
\;=\;
\frac{1}{6}
\sum\limits_{i,\,j,\,k = 1}^{n}
\left(\mbox{\ae}_{12}^{ij}\,\mbox{\ae}_{16}^{ik}  \;-\; 
\mbox{\ae}_{16}^{ij}\,\mbox{\ae}_{12}^{ik} \right)^2 ,
\label{binC_3}
\end{eqnarray}

\noindent
where $\,\mbox{\ae}_{52}^{ij}\,$ (in the framework of the usual 
6 by 6 matrix formalism for the linear beam dynamics) is the coefficient 
that connects variation of the particle path length 
with variation of the particle transverse momentum and which can be expressed 
using elements of the matrix $\,B_{ij}\,$ as follows

\noindent
\begin{eqnarray}
\mbox{\ae}_{52}^{ij} \;=\;
\mbox{\ae}_{22}^{ij} \,\mbox{\ae}_{16}^{ij} \;-\; 
\mbox{\ae}_{12}^{ij} \,\mbox{\ae}_{26}^{ij}\,.
\label{binC_4}
\end{eqnarray}

From (\ref{binC_3}) one sees, that the matrix $\,M\,$ will have
the full column rank if and only if
there are at least three beam position monitors with indices
$\,i,\,j\,$ and $\,k\,$ such that the transport
matrices between them satisfy the condition

\noindent
\begin{eqnarray}
\mbox{\ae}_{52}^{ij} \,\mbox{\ae}_{12}^{jk} \;-\; 
\mbox{\ae}_{12}^{ij} \,\mbox{\ae}_{16}^{jk} \;\neq\; 0
\label{binC_3_0}
\end{eqnarray}

\noindent
or (equivalently) the condition

\noindent
\begin{eqnarray}
\mbox{\ae}_{12}^{ij}\,\mbox{\ae}_{16}^{ik}  \;-\; 
\mbox{\ae}_{16}^{ij}\,\mbox{\ae}_{12}^{ik} \;\neq\; 0 .
\label{binC_3_1}
\end{eqnarray}

Note that both conditions, (\ref{binC_3_0}) and (\ref{binC_3_1}), 
involve elements of two transfer matrices, but
while (\ref{binC_3_0}) uses matrices between neighboring 
BPMs ($B_{ij}\,$ and $\,B_{jk}$), condition (\ref{binC_3_1}) operates 
with the transport matrices from
first to two remaining BPMs ($B_{ij}\,$ and $\,B_{ik}$).  
In simple words the condition (\ref{binC_3_1}), for example, means that
one can not vary particle transverse momentum and particle energy at the
first BPM location in such a fashion that these variations are invisible at
the two downstream BPMs.

\section{Beam Dynamical Parametrization of\\
Covariance Matrix of Reconstruction Errors}

\hspace{0.5cm}
Let $\,A(r_1, \, r_2)\,$ be a matrix which transport particle coordinates
from the point with the longitudinal position  $\,s = r_1\,$ to the point
with the longitudinal position  $\,s = r_2\,$

\noindent
\begin{eqnarray}
A(r_1,\,r_2) \, = \,
\left(
\begin{array}{lll}
m_{11} & m_{12} & m_{16}\\
m_{21} & m_{22} & m_{26}\\
0 & 0 & 1
\end{array}
\right),
\hspace{0.5cm}
m_{11} \, m_{22} - m_{12} \, m_{21} \, = \, 1.
\label{bdp_1}
\end{eqnarray}

\noindent
Similar to \cite{ThePaper}, one can easily show that for 
any given value of $\,\vec{\,\varsigma}\,$
the estimate of the difference orbit parameters  
$\,\delta \vec{z}_{\varsigma}\,$
propagates along the beam line exactly as particle trajectory

\noindent
\begin{eqnarray}
\delta \vec{z}_{\varsigma}(r_2) \; = \;
A( r_1 , \, r_2 ) \cdot \delta \vec{z}_{\varsigma}(r_1),
\label{bdp_2}
\end{eqnarray}

\noindent
as one changes the position of the reconstruction point.
So again we can consider a virtual beam consisting from 
virtual particles obtained as a result of application of
formula (\ref{SEC1_15}) to ``all possible values'' of the
error vector $\,\vec{\,\varsigma}$. The dynamics of the
centroid of this beam $\,\delta \vec{z}_0\,$ coincides with the dynamics
of the true difference orbit 

\noindent
\begin{eqnarray}
\delta \vec{z}_0(r) \;\stackrel{\rm def}{=}\;
\big<\delta \vec{z}_{\varsigma}(r)\big>\;=\;
\delta \vec{z}(r) \,,
\label{bdp_2_0}
\end{eqnarray}

\noindent
and the error covariance matrix (\ref{SEC1_16}) can be treated
as the matrix of the second central moments of this virtual beam
distribution and satisfies the usual transport equation

\noindent
\begin{eqnarray}
V_z(r_2) \;=\; A( r_1 , \, r_2 ) \, V_z(r_1) \,
A^{\top}( r_1 , \, r_2 ) .
\label{bdp_3}
\end{eqnarray}

Consequently, for the description of the propagation of
the reconstruction errors along the beam line,
one can use the accelerator physics notations
and represent the error covariance matrix in the familiar form

\noindent
\begin{eqnarray}
V_z =
\left( M_{\varsigma}^{\top} M_{\varsigma} \right)^{-1} =
\epsilon_{\varsigma} 
\, 
\left(
\begin{array}{rrr}
  \beta_{\varsigma} & -\alpha_{\varsigma} & 0 \\
-\alpha_{\varsigma} &  \gamma_{\varsigma} & 0 \\
 0 & 0 & 0
\end{array}
\right)
+ 
\Delta_{\varsigma}^2 
\,
\left(
\begin{array}{c}
\eta_{x, \varsigma} \\
\eta_{p, \varsigma} \\
1
\end{array}
\right)
\left(
\begin{array}{c}
\eta_{x, \varsigma} \\
\eta_{p, \varsigma} \\
1
\end{array}
\right)^{\top} \;=
\nonumber
\end{eqnarray}

\noindent
\begin{eqnarray}
=\;
\left(
\begin{array}{ccc}
 \epsilon_{\varsigma} \,\beta_{\varsigma} + \Delta_{\varsigma}^2  \,\eta_{x, \varsigma}^2 & 
-\epsilon_{\varsigma} \, \alpha_{\varsigma} + \Delta_{\varsigma}^2  \,\eta_{x, \varsigma} \,
\eta_{p, \varsigma} & 
\Delta_{\varsigma}^2 \, \eta_{x, \varsigma} \\
-\epsilon_{\varsigma}  \,\alpha_{\varsigma} + \Delta_{\varsigma}^2  \,
\eta_{x, \varsigma} \,\eta_{p, \varsigma} & 
 \epsilon_{\varsigma}  \,\gamma_{\varsigma} + \Delta_{\varsigma}^2 \, \eta_{p, \varsigma}^2 & 
\Delta_{\varsigma}^2 \, \eta_{p, \varsigma} \\
\Delta_{\varsigma}^2  \,\eta_{x, \varsigma} &
\Delta_{\varsigma}^2  \,\eta_{p, \varsigma} &
\Delta_{\varsigma}^2
\end{array}
\right).
\label{TwCVM_1}
\end{eqnarray}

As usual for the particle dynamics, this parametrization has two invariants
(quantities which are independent from the position of the reconstruction point),
namely transverse error emittance $\,\epsilon_{\varsigma}\,$
and rms error energy spread $\,\Delta_{\varsigma}$,
which can be calculated according to the formulas

\noindent
\begin{eqnarray}
\epsilon_{\varsigma} \;=\;
\frac{1}{\sqrt{\Gamma\left(\vec{a}_{\varsigma}, \, \vec{c}_{\varsigma}\right)}},
\hspace{1.0cm}
\Delta_{\varsigma} \;=\;
\sqrt{\frac{\Gamma\left(\vec{a}_{\varsigma}, \, \vec{c}_{\varsigma}\right)}
{\Gamma\left(\vec{a}_{\varsigma}, \, \vec{c}_{\varsigma}, \, \vec{g}_{\varsigma}\right)}},
\label{TwCVM_2}
\end{eqnarray}

\noindent
where we have used the notations

\noindent
\begin{eqnarray}
\vec{a}_{\varsigma} \;=\; R_{\varsigma}^{-\top} \vec{a}, 
\hspace{0.7cm}
\vec{c}_{\varsigma} \;=\; R_{\varsigma}^{-\top} \vec{c}, 
\hspace{0.7cm}
\vec{g}_{\varsigma} \;=\; R_{\varsigma}^{-\top} \vec{g}
\label{m3}
\end{eqnarray}

\noindent
and $\,\Gamma\left(\vec{u}_1,\,\ldots\,\vec{u}_m\right)\,$ is the
Gram determinant of the vectors $\,\vec{u}_1,\,\ldots\,\vec{u}_m$.

The error Twiss parameters, of course, remain the same as they 
were earlier published in \cite{ThePaper}, namely 

\noindent
\begin{eqnarray}
\beta_{\varsigma}(r) =
\epsilon_{\varsigma} \,
\left\| \vec{c}_{\varsigma}(r)\right\|_2^2,
\hspace{0.35cm}
\alpha_{\varsigma}(r) =
\epsilon_{\varsigma} \,
\left( \vec{a}_{\varsigma}(r) \cdot \vec{c}_{\varsigma}(r)\right),
\hspace{0.35cm}
\gamma_{\varsigma}(r) =
\epsilon_{\varsigma} \,
\left\| \vec{a}_{\varsigma}(r)\right\|_2^2,
\label{TwCVM_3}
\end{eqnarray}

\noindent
and for the new objects, the coordinate and momentum error dispersions, we have

\noindent
\begin{eqnarray}
\eta_{x, \varsigma}(r) \;=\;
\epsilon_{\varsigma} \,
\Big(
\alpha_{\varsigma}(r) \,
\left( \vec{c}_{\varsigma}(r) \cdot \vec{g}_{\varsigma}(r)\right)
\;-\;
\beta_{\varsigma}(r) \,
\left( \vec{a}_{\varsigma}(r) \cdot \vec{g}_{\varsigma}(r)\right)
\Big),
\label{TwCVM_4}
\end{eqnarray}

\noindent
\begin{eqnarray}
\eta_{p, \varsigma}(r) \;=\;
\epsilon_{\varsigma} \,
\Big(
\alpha_{\varsigma}(r) \,
\left( \vec{a}_{\varsigma}(r) \cdot \vec{g}_{\varsigma}(r)\right)
\;-\;
\gamma_{\varsigma}(r) \,
\left( \vec{c}_{\varsigma}(r) \cdot \vec{g}_{\varsigma}(r)\right)
\Big).
\label{TwCVM_5}
\end{eqnarray}

As it was shown in \cite{ThePaper}, the error Twiss parameters (\ref{TwCVM_3}) 
are not simply one of many betatron functions which could propagate
through our beam line, they are by themselves solutions of some minimization
problem and are uniquely determined by transport matrices between BPM
locations and by BPM resolutions. And we would like to show, that
the same is true also for the error dispersions (\ref{TwCVM_4}) and (\ref{TwCVM_5}).

Let $\,\eta_x(r)\,$ and $\,\eta_p(r)\,$ be some dispersions specified
in the reconstruction point. Then the corresponding coordinate dispersion
at the $m$-th BPM location can be calculated as follows

\noindent
\begin{eqnarray}
\eta_x(s_m) \;=\; a_m(r) \,\eta_x(r) \,+\, c_m(r) \,\eta_p(r) \,+\, g_m(r) .
\label{d1}
\end{eqnarray}

\noindent
Consider a vector

\noindent
\begin{eqnarray}
\vec{D}\left(r, \,\eta_x(r),\, \eta_p(r)\right) 
= R_{\varsigma}^{-\top} \left(\eta_x(s_1) , \ldots , \eta_x(s_n)\right)^{\top}
= \eta_x(r) \,\vec{a}_{\varsigma}
+ \eta_p(r) \,\vec{c}_{\varsigma}
+ \vec{g}_{\varsigma} 
\label{d2}
\end{eqnarray}

\noindent
and a minimization problem

\noindent
\begin{eqnarray}
\min_{\eta_x(r), \, \eta_p(r)}
\big\|\, \vec{D}\left(r, \,\eta_x(r), \,\eta_p(r)\right)\,\big\|_2^2 .
\label{d3}
\end{eqnarray}

By standard means it is not difficult to show that if
$\,\Gamma\left(\vec{a}, \, \vec{c}\right) \neq 0\,$ then
the solution of this minimization problem is unique and 
is given by the formulas (\ref{TwCVM_4}) and (\ref{TwCVM_5}).

If, additionally, $\,\Gamma\left(\vec{a}, \, \vec{c}, \, \vec{g}\right) \neq 0\,$
then the minimum in (\ref{d3}) is bigger than zero 
(and is equal to zero otherwise) and the following identity holds

\noindent
\begin{eqnarray}
\big\|\, \vec{D}\left(r, \,\eta_{x,\varsigma}(r),\, \eta_{p,\varsigma}(r)\right)\,\big\|_2^2
\;=\; \frac{1}{\Delta_{\varsigma}^2}\,.
\label{d4}
\end{eqnarray}

Note that geometrically the vector
$\,\eta_x(r) \,\vec{a}_{\varsigma} + \eta_p(r) \,\vec{c}_{\varsigma}\,$
is nothing else as
taken with an opposite sign projection of the
vector $\,\vec{g}_{\varsigma}\,$ onto a linear subspace 
formed by vectors $\,\vec{a}_{\varsigma}\,$ and $\,\vec{c}_{\varsigma}$
and hence the vector 
$\,\vec{D}\left(r, \,\eta_{x,\varsigma}(r),\, \eta_{p,\varsigma}(r)\right)\,$ 
is orthogonal to both, vector $\,\vec{a}_{\varsigma}\,$ and 
vector $\,\vec{c}_{\varsigma}$. 

To finish this section let us, for the case when readings of different BPMs  
are uncorrelated, i.e. when the covariance matrix $\,V_{\varsigma}\,$
is a positive diagonal matrix 

\noindent
\begin{eqnarray}
V_{\varsigma} \;=\;
\mbox{diag}
\left(\,
\sigma_1^2 ,\, \sigma_2^2 ,\, \ldots ,\, \sigma_n^2\,
\right)\;>\; 0\,,
\label{SEC1_17}
\end{eqnarray}

\noindent
write down the following useful expressions for the Gram determinants

\noindent
\begin{eqnarray}
\Gamma\left(\vec{a}_{\varsigma},\,\vec{c}_{\varsigma}\right) 
=
\frac{1}{2}
\sum\limits_{i,\,j = 1}^{n}
\left(\frac{\mbox{\ae}_{12}^{ij}}{\sigma_i \,\sigma_j}\right)^2,
\label{binC_29}
\end{eqnarray}

\noindent
\begin{eqnarray}
\Gamma\left(\vec{a}_{\varsigma},\vec{c}_{\varsigma},\vec{g}_{\varsigma}\right) 
=
\frac{1}{6}
\sum\limits_{i,\,j,\,k = 1}^{n}
\left(\frac{\mbox{\ae}_{52}^{ij} \mbox{\ae}_{12}^{jk} - 
\mbox{\ae}_{12}^{ij} \mbox{\ae}_{16}^{jk}}{\sigma_i \,\sigma_j\, \sigma_k}\right)^2
=
\frac{1}{6}
\sum\limits_{i,\,j,\,k = 1}^{n}
\left(\frac{\mbox{\ae}_{12}^{ij} \mbox{\ae}_{16}^{ik} - 
\mbox{\ae}_{16}^{ij} \mbox{\ae}_{12}^{ik}}{\sigma_i \,\sigma_j \,\sigma_k} \right)^2 ,
\label{binC_30}
\end{eqnarray}

\noindent
which enter formulas (\ref{TwCVM_2}) for the transverse error emittance 
and for the rms error energy spread.

\section{Three BPMs in Symmetric Arrangement}

\hspace{0.5cm}
Let us assume that we have three beam position monitors in our
beam line which deliver uncorrelated readings with rms
resolutions $\sigma_1$, $\sigma_2$ and $\sigma_3$, and
let $\,B_{12}\,$ and $\,B_{23}\,$ be transfer matrices between 
first and second, and between second and third BPM locations respectively

\noindent
\begin{eqnarray}
B_{12}\;=\;
\left(
\begin{array}{ccc}
r_{11} & r_{12}& r_{16}\\
r_{21} & r_{22}& r_{26}\\
0 & 0 & 1
\end{array}
\right),
\hspace{0.5cm}
B_{23}\;=\;
\left(
\begin{array}{ccc}
m_{11} & m_{12}& m_{16}\\
m_{21} & m_{22}& m_{26}\\
0 & 0 & 1
\end{array}
\right).
\label{SEC_ThreeBPM_0}
\end{eqnarray}

When the phase advance between the first and the second BPMs or
the phase advance between the second and the third BPMs is not multiple
of $\,180^{\circ}\,$, i.e. when

\noindent
\begin{eqnarray}
r_{12}^2 \,+\, m_{12}^2 \,\neq\,0,
\label{SEC_ThreeBPM_001}
\end{eqnarray}

\noindent
this system can be used for the measurement of the transverse
orbit jitter with the transverse error emittance given by
the following expression

\noindent
\begin{eqnarray}
\epsilon_{\varsigma} \;=\;
\frac{\sigma_1 \,\sigma_2\, \sigma_3}
{\sqrt{\,\sigma_1^2 \,m_{12}^2 \,+\,
\sigma_2^2\,\left(m_{11}\,r_{12}\,+\,r_{22}\,m_{12}\right)^2 \,+\, \sigma_3^2\, r_{12}^2\,}}\,.
\label{SEC_ThreeBPM_002}
\end{eqnarray}

In order to be able to resolve both, transverse and energy, jitters simultaneously
we have to assume, additionally to (\ref{SEC_ThreeBPM_001}), that

\noindent
\begin{eqnarray}
m_{12}\,r_{52}\,-\,r_{12}\,m_{16} \,\neq\,0\,,
\label{SEC_ThreeBPM_0021}
\end{eqnarray}

\noindent
where the $\,r_{52}\,$  and $\,r_{51}\,$  coefficients
can be expressed using elements of the matrix $\,B_{12}\,$ as follows

\noindent
\begin{eqnarray}
\left\{
\begin{array}{ccc}
r_{51} &=& r_{21} r_{16} - r_{11} r_{26}\\
r_{52} &=& r_{22} r_{16} - r_{12} r_{26}
\end{array}
\right. 
.
\label{SEC_ThreeBPM_1_1}
\end{eqnarray}

\noindent
With (\ref{SEC_ThreeBPM_001}) and (\ref{SEC_ThreeBPM_0021}) satisfied,
we obtain for the square of the rms error energy spread 

\noindent
\begin{eqnarray}
\Delta^2_{\varsigma} \;=\;
\frac{\,\sigma_1^2 \,m_{12}^2 \,+\,
\sigma_2^2\,\left(m_{11}\,r_{12}\,+\,r_{22}\,m_{12}\right)^2 \,+\, \sigma_3^2\, r_{12}^2\,}
{\left(m_{12}\,r_{52}\,-\,r_{12}\,m_{16}\right)^2}\,.
\label{SEC_ThreeBPM_003}
\end{eqnarray}

To complete description of the covariance matrix of the reconstruction errors
(\ref{TwCVM_1}) for the three BPM case,
we also need formulas for the error coordinate and momentum dispersions, and
for the error betatron functions. And although it is not very difficult
to provide some formulas using (\ref{TwCVM_3}), (\ref{TwCVM_4}) and (\ref{TwCVM_5}), 
the results are not very informative and it is not easy to derive some nontrivial
conclusions from them. 
So in this section we will give more digestible
expressions for error dispersions and error betatron 
functions making an additional simplifying assumption about our measurement system
that the transfer matrix $\,B_{23}\,$ 
between the second and the third BPM is not an arbitrary beam transport matrix, but 
is obtained as a result of some symmetry manipulation with the transfer
matrix between the first and the second BPM. 

\subsection{Mirror Symmetric Optical System}

\hspace{0.5cm}
Let a magnet system between the second and the third BPMs be a mirror symmetric
image of the magnet structure between the first and the second 
BPM locations. Then 

\noindent
\begin{eqnarray}
B_{23}\;=\;
\left(
\begin{array}{ccc}
r_{22} & r_{12}& -r_{52}\\
r_{21} & r_{11}& -r_{51}\\
0 & 0 & 1
\end{array}
\right) .
\label{SEC_ThreeBPM_01}
\end{eqnarray}

\noindent
The transverse error emittance of this measurement system is given by

\noindent
\begin{eqnarray}
\epsilon_{\varsigma} \;=\;
\frac{1}{\left| r_{12}\right|} \cdot
\frac{\sigma_1 \,\sigma_2\, \sigma_3}
{\sqrt{\sigma_1^2 \,+\, 4 \,\sigma_2^2\,r_{22}^2 \,+\, \sigma_3^2}}\,,
\label{SEC_ThreeBPM_1}
\end{eqnarray}

\noindent
and the error betatron functions at the BPM locations can be calculated as follows

\noindent
\begin{eqnarray}
\beta_{\varsigma}(s_1) \;=\; 
\left| r_{12} \right| \cdot
\frac{\sigma_1}{\sigma_2 \, \sigma_3} \cdot
\frac{ 4 \,\sigma_2^2\,r_{22}^2 \,+\, \sigma_3^2}
{\sqrt{\sigma_1^2 \,+\, 4 \,\sigma_2^2\,r_{22}^2 \,+\, \sigma_3^2}} ,
\label{SEC_ThreeBPM_2}
\end{eqnarray}

\noindent
\begin{eqnarray}
\alpha_{\varsigma}(s_1) \;=\;
\mbox{sign}\left(r_{12}\right)\cdot
\frac{\sigma_1}{\sigma_2 \, \sigma_3} \cdot
\frac{\, 2 \,\sigma_2^2\,r_{22} \,\left(1 \,+\, 2 r_{12} r_{21} \right) \,+\, \sigma_3^2 \,r_{11}\,}
{\sqrt{\sigma_1^2 \,+\, 4 \,\sigma_2^2\,r_{22}^2 \,+\, \sigma_3^2}} ,
\label{SEC_ThreeBPM_3}
\end{eqnarray}

\noindent
\begin{eqnarray}
\beta_{\varsigma}(s_2) \;=\;
\left| r_{12} \right| \cdot
\frac{\sigma_2}{\sigma_1 \, \sigma_3} \cdot
\frac{\,\sigma_1^2 \,+\, \sigma_3^2\,}
{\sqrt{\sigma_1^2 \,+\, 4 \,\sigma_2^2\,r_{22}^2 \,+\, \sigma_3^2}} ,
\label{SEC_ThreeBPM_4}
\end{eqnarray}

\noindent
\begin{eqnarray}
\alpha_{\varsigma}(s_2) \;=\;
\mbox{sign}\left(r_{12}\right)\cdot
\left(\frac{\sigma_1}{\sigma_3} -  \frac{\sigma_3}{\sigma_1}\right) \cdot
\frac{\, \sigma_2 \,r_{22} \,}
{\sqrt{\sigma_1^2 \,+\, 4 \,\sigma_2^2\,r_{22}^2 \,+\, \sigma_3^2}} ,
\label{SEC_ThreeBPM_4_1}
\end{eqnarray}

\noindent
\begin{eqnarray}
\beta_{\varsigma}(s_3) \;=\; 
\left| r_{12} \right| \cdot
\frac{\sigma_3}{\sigma_1 \, \sigma_2} \cdot
\frac{ \sigma_1^2 \,+\,4 \,\sigma_2^2\,r_{22}^2 }
{\sqrt{\sigma_1^2 \,+\, 4 \,\sigma_2^2\,r_{22}^2 \,+\, \sigma_3^2}} ,
\label{SEC_ThreeBPM_2_3}
\end{eqnarray}

\noindent
\begin{eqnarray}
\alpha_{\varsigma}(s_3) \;=\;
-\mbox{sign}\left(r_{12}\right)\cdot
\frac{\sigma_3}{\sigma_1 \, \sigma_2} \cdot
\frac{\,\sigma_1^2 \,r_{11} \,+\,2 \,\sigma_2^2\,r_{22} \,\left(1 \,+\, 2 r_{12} r_{21} \right)\,}
{\sqrt{\sigma_1^2 \,+\, 4 \,\sigma_2^2\,r_{22}^2 \,+\, \sigma_3^2}} .
\label{SEC_ThreeBPM_3_3}
\end{eqnarray}

If we will assume that BPM resolutions follow mirror symmetry of the system,
which means that $\,\sigma_1\,$ is equal to $\,\sigma_3$, then,
as it could be expected, the error Twiss parameters will satisfy
the following symmetry relations

\noindent
\begin{eqnarray}
\beta_{\varsigma}(s_3) = \beta_{\varsigma}(s_1),
\hspace{0.7cm}
\alpha_{\varsigma}(s_3) = -\alpha_{\varsigma}(s_1),
\hspace{0.7cm}
\alpha_{\varsigma}(s_2) = 0.
\label{SEC_ThreeBPM_BS}
\end{eqnarray}

For the square of the error energy spread we have the following
expression 

\noindent
\begin{eqnarray}
\Delta^2_{\varsigma} \;=\;
\frac{\sigma_1^2 \,+\, 4 \,\sigma_2^2\,r_{22}^2 \,+\, \sigma_3^2}
{4 \, r_{52}^2}\,,
\label{SEC_ThreeBPM_1_0}
\end{eqnarray}

\noindent
and the coordinate and momentum error dispersions 
at the BPM locations are given below

\noindent
\begin{eqnarray}
\eta_{x, \varsigma}(s_1) \;=\;
-\frac{ 2 \, \sigma_1^2 \, r_{52}}
{\sigma_1^2 \,+\, 4 \,\sigma_2^2\,r_{22}^2 \,+\, \sigma_3^2} ,
\label{SEC_ThreeBPM_5}
\end{eqnarray}

\noindent
\begin{eqnarray}
\eta_{p, \varsigma}(s_1) \;=\;
\frac{\,\sigma_1^2 \,\left( r_{16} \,+\, 2 \,r_{12}\, r_{51}\right) \,-\,
 4 \,\sigma_2^2 \,r_{12}\,r_{22}\,r_{26} 
\,-\, \sigma_3^2\, r_{16}\,}
{r_{12} \cdot \left(\sigma_1^2 \,+\, 4 \,\sigma_2^2\,r_{22}^2 \,+\, \sigma_3^2\right)} ,
\label{SEC_ThreeBPM_6}
\end{eqnarray}

\noindent
\begin{eqnarray}
\eta_{x, \varsigma}(s_2) \;=\;
\frac{ 4 \, \sigma_2^2 \, r_{22}\, r_{52}}
{\sigma_1^2 \,+\, 4 \,\sigma_2^2\,r_{22}^2 \,+\, \sigma_3^2} ,
\label{SEC_ThreeBPM_5_2}
\end{eqnarray}

\noindent
\begin{eqnarray}
\eta_{p, \varsigma}(s_2) \;=\;
\left(\sigma_1^2 \,-\,\sigma_3^2 \right) \cdot
\frac{r_{52}}
{r_{12} \cdot \left(\sigma_1^2 \,+\, 4 \,\sigma_2^2\,r_{22}^2 \,+\, \sigma_3^2\right)} ,
\label{SEC_ThreeBPM_6_2}
\end{eqnarray}

\noindent
\begin{eqnarray}
\eta_{x, \varsigma}(s_3) \;=\;
-\frac{ 2 \, \sigma_3^2 \, r_{52}}
{\sigma_1^2 \,+\, 4 \,\sigma_2^2\,r_{22}^2 \,+\, \sigma_3^2} ,
\label{SEC_ThreeBPM_5_3}
\end{eqnarray}

\noindent
\begin{eqnarray}
\eta_{p, \varsigma}(s_3) \;=\;
\frac{\,\sigma_1^2\, r_{16}\,+\,
 4 \,\sigma_2^2 \,r_{12}\,r_{22}\,r_{26} 
\,-\,
\sigma_3^2 \,\left( r_{16} \,+\, 2 \,r_{12}\, r_{51}\right) \,}
{r_{12} \cdot \left(\sigma_1^2 \,+\, 4 \,\sigma_2^2\,r_{22}^2 \,+\, \sigma_3^2\right)} .
\label{SEC_ThreeBPM_6_3}
\end{eqnarray}

One sees that if BPM resolutions will follow mirror symmetry of the system,
they, similar to the error betatron functions, will satisfy

\noindent
\begin{eqnarray}
\eta_{x, \varsigma}(s_3) = \eta_{x, \varsigma}(s_1),
\hspace{0.7cm}
\eta_{p, \varsigma}(s_3) = -\eta_{p, \varsigma}(s_1),
\hspace{0.7cm}
\eta_{p, \varsigma}(s_2) = 0,
\label{SEC_ThreeBPM_DS}
\end{eqnarray}

\noindent
independently if $\,r_{26}\,$ is equal to 
zero or not.\footnote{Let us remind, that the condition $\,r_{26} = 0\,$
applied in the symmetry point 
is the necessary and sufficient condition for 
the total transfer matrix of the mirror symmetric system to be achromatic.}
One also sees that if mirror symmetric system can be used for energy
jitter measurement (i.e. if $\,r_{52} \neq 0\,$), then the error
dispersion is nonzero at the system entrance and exit, again
independently if $\,r_{26}\,$ is equal to 
zero or not. 

\begin{figure} 
\centerline{\hbox{\psfig{figure=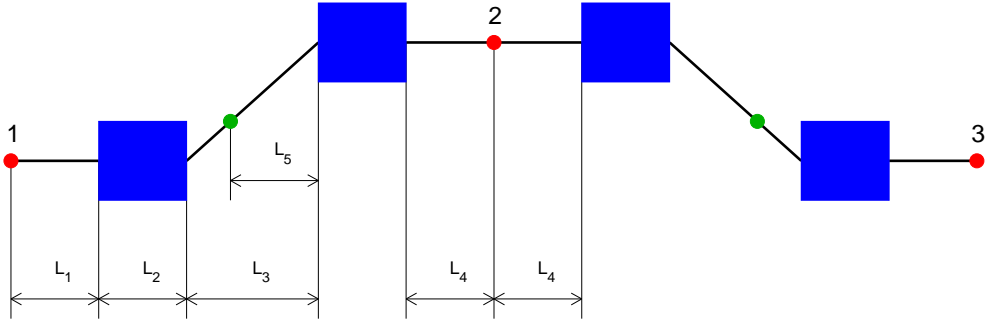,width=14.0cm}}}
\caption{Schematic layout of four bend chicane.}
\label{fig_P2_0}
\end{figure}

As a more specific example, let us consider three BPMs integrated into four bend
magnetic chicane, as shown by red circles in figure 1. For this system

\noindent
\begin{eqnarray}
B_{12}\;=\;
\left(
\begin{array}{ccc}
1 & r_{12}& r_{16}\\
0 & 1 & 0\\
0 & 0 & 1
\end{array}
\right),
\label{Chicane_1}
\end{eqnarray}

\noindent
where

\noindent
\begin{eqnarray}
r_{12} \,=\, L_1 \,+\, L_4 \,+\, \frac{2\, L_2}{\cos(\varphi)} \,+\, \frac{L_3}{\cos^3(\varphi)}
\, \neq\, 0
\label{Chicane_2}
\end{eqnarray}

\noindent
and

\noindent
\begin{eqnarray}
r_{16} \,=\, r_{52} \,=\, 
\frac{1}{\cos(\varphi)} \cdot
\left(
2 \,L_2\, \tan(\varphi / 2)
\,+\, \frac{L_3}{\cos(\varphi)}\, \tan(\varphi)
\right)
\, \neq\, 0.
\label{Chicane_3}
\end{eqnarray}

\noindent
Therefore this system always can be used for the transverse and
energy jitter measurement, and, as a concrete case, 
let us consider the first bunch compressor 
of the FLASH facility \cite{FLASH_1, FLASH_2}, 
which is the four bend chicane of the discussed layout. 
The typical deflection angle for this chicane is about $18^{\circ}$ 
and the distances $L_2$ and $L_3$ are equal each other and are equal to $0.5\,m$
(see, for example, \cite{PedroBC2}).
Let us assume that the first and the third BPMs (orbit BPMs) have the same
rms resolutions $\,\sigma_1 = \sigma_3 = \sigma_{orb}\,$ and for the second BPM
(energy BPM) let us introduce the notation $\,\sigma_2 = \sigma_{enr}$. 
Let $\,\Delta_{des}\,$ will be energy jitter resolution desired for the
system. With these numbers and notations, and using the usual three sigma criterion
$\,(3 \Delta_{\varsigma} \leq \Delta_{des})\,$
we obtain from (\ref{SEC_ThreeBPM_1_0}) the following inequality 

\noindent
\begin{eqnarray}
\sigma_{orb}^2 \,+\, 2 \,\sigma_{enr}^2 \;\leq\;
0.02663 \cdot \Delta_{des}^2\,,
\label{Chicane_4}
\end{eqnarray}

\noindent
which gives us limitations on the range of the BPM resolutions which can provide
the required precision for the energy jitter measurement. Figure 2,
for example,
shows the area of acceptable BPM resolutions defined by the inequality (\ref{Chicane_4}) for
$\,\Delta_{des} = 5\cdot10^{-5}$.

\begin{figure} 
\centerline{\hbox{\psfig{figure=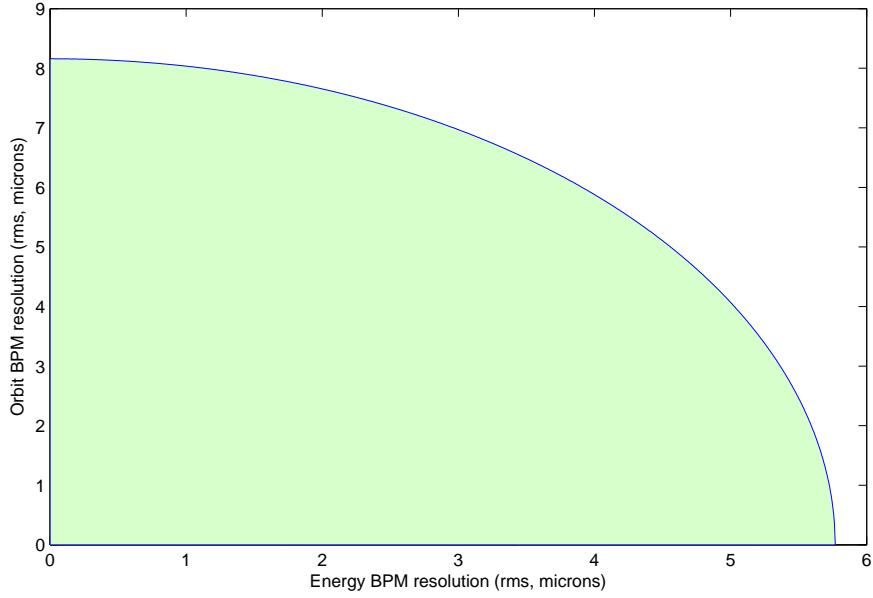,width=14.0cm}}}
\caption{Resolutions of orbit and energy BPMs (shaded area)  
which are needed in order to be able to resolve energy jitter $5\cdot 10^{-5}$ in the first
bunch compressor of the FLASH facility. BPMs are positioned as shown by
red circles in figure 1.}
\label{fig_P2_4}
\end{figure}

To finish the chicane discussion,
let us move the first and the third BPMs into positions shown as
green circles in figure 1. For this case

\noindent
\begin{eqnarray}
B_{12}\;=\;
\left(
\begin{array}{ccc}
r_{11} & r_{12}& r_{16}\\
0 & r_{22} & r_{26}\\
0 & 0 & 1
\end{array}
\right),
\label{Chicane_5}
\end{eqnarray}

\noindent
\begin{eqnarray}
r_{22} \,=\,\frac{1}{r_{11}} \,=\,\cos(\varphi),
\hspace{1.0cm}
r_{12} \,=\,
\frac{L_5}{\cos^2(\varphi)} \,+\,L_2 \,+\,L_4 \,\cos(\varphi), 
\label{Chicane_6}
\end{eqnarray}

\noindent
\begin{eqnarray}
r_{16} \,=\,- L_2 \,\tan(\varphi / 2) \,-\,L_4 \,\sin(\varphi),
\hspace{1.0cm}
r_{26} \,=\, -\sin(\varphi), 
\label{Chicane_7}
\end{eqnarray}

\noindent
\begin{eqnarray}
r_{52} \,=\,L_2 \,\tan(\varphi / 2) \,+\,\frac{L_5}{\cos(\varphi)} \,\tan(\varphi),
\label{Chicane_8}
\end{eqnarray}

\noindent
and one sees that this BPM positioning still can be used
for the jitter measurement, because $\,r_{12} \neq 0\,$
and $\,r_{52} \neq 0$, but both, the transverse error emittance
and the error energy spread become larger 
(for the same BPM resolutions) than for the original BPM layout.
Nevertheless, it is a good example of a mirror symmetric system
for which  $\,r_{52} \,\neq\, r_{16}\,$ and the total transfer
matrix is not achromatic. 

\subsection{Mirror Antisymmetric Case}

\hspace{0.5cm}
If a magnet system between the second and the third BPMs is a mirror 
antisymmetric image of the first part of the system, then 

\noindent
\begin{eqnarray}
B_{23}\;=\;
\left(
\begin{array}{ccc}
r_{22} & r_{12}& r_{52}\\
r_{21} & r_{11}& r_{51}\\
0 & 0 & 1
\end{array}
\right) .
\label{MAC_01}
\end{eqnarray}

The transverse error emittance and the error beta
functions remain the same as for the mirror symmetric case, but
the measurement of the energy jitter is not possible anymore, independent of
the BPM resolutions following symmetry of the system or not. The coordinate
error dispersion is always zero at the BPM locations with the
momentum error dispersion taking the values 

\noindent
\begin{eqnarray}
\eta_{p, \varsigma}(s_1) \,=\, \eta_{p, \varsigma}(s_3) \,=\, -\frac{r_{16}}{r_{12}},
\hspace{1.0cm}
\eta_{p, \varsigma}(s_2) \,=\,  -\frac{r_{52}}{r_{12}} ,
\label{MAC_02}
\end{eqnarray}

\noindent
which are independent from BPM resolutions.
Note that this impossibility of the energy jitter measurement
does not depend on the value of $\,r_{16}\,$ which could be zero
or not.\footnote{The condition $\,r_{16} = 0\,$
is the necessary and sufficient condition for 
the total transfer matrix of the mirror antisymmetric 
system to be achromatic.}

\subsection{Two Periodic System}

\hspace{0.5cm}
Let us assume that our measurement system is periodic,
by which we mean that $\,B_{23} = B_{12}$. We named it in the 
title as two periodic owing the fact that two equal transfer matrices
are involved, but, more correctly, it should be treated as
a three cell system because we consider three BPMs. Note that 
general periodic measurement systems constructed from $n$ identical cells will
be studied in details in the next section, but with additional simplifying assumptions
that the cell transfer matrix allows periodic beam transport and that all BPMs
have the same resolutions.

The transverse error emittance of the two periodic system can be expressed in the form

\noindent
\begin{eqnarray}
\epsilon_{\varsigma} \;=\;
\frac{1}{\left| r_{12}\right|} \cdot
\frac{\sigma_1 \,\sigma_2\, \sigma_3}
{\sqrt{\sigma_1^2 \,+\, \sigma_2^2\cdot\mbox{tr}_x^2\left(B_{12}\right) \,+\, \sigma_3^2}}\,,
\label{TPS_01}
\end{eqnarray}

\noindent
where

\noindent
\begin{eqnarray}
\mbox{tr}_x\left(B_{12}\right) \;=\; r_{11}\,+\,r_{22}\,,
\label{TPS_01_D}
\end{eqnarray}

\noindent
and calculation of the error betatron functions gives the following results

\noindent
\begin{eqnarray}
\beta_{\varsigma}(s_1) \;=\; 
\left| r_{12} \right| \cdot
\frac{\sigma_1}{\sigma_2 \, \sigma_3} \cdot
\frac{ \sigma_2^2\cdot\mbox{tr}_x^2\left(B_{12}\right) \,+\, \sigma_3^2}
{\sqrt{\sigma_1^2 \,+\, \sigma_2^2\cdot\mbox{tr}_x^2\left(B_{12}\right) \,+\, \sigma_3^2}}\,,
\label{TPS_01_1}
\end{eqnarray}

\noindent
\begin{eqnarray}
\alpha_{\varsigma}(s_1) \;=\;
\mbox{sign}\left(r_{12}\right)\cdot
\frac{\sigma_1}{\sigma_2 \, \sigma_3} \cdot
\frac{\sigma_2^2\cdot\mbox{tr}_x\left(B_{12}\right)\cdot\left(r_{11}\cdot\mbox{tr}_x\left(B_{12}\right) \,-\,1\right) \,+\, \sigma_3^2 \,r_{11}\,}
{\sqrt{\sigma_1^2 \,+\, \sigma_2^2\cdot\mbox{tr}_x^2\left(B_{12}\right) \,+\, \sigma_3^2}}\,,
\label{TPS_01_2}
\end{eqnarray}

\noindent
\begin{eqnarray}
\beta_{\varsigma}(s_2) \;=\; 
\left| r_{12} \right| \cdot
\frac{\sigma_2}{\sigma_1 \, \sigma_3} \cdot
\frac{ \sigma_1^2 \,+\, \sigma_3^2}
{\sqrt{\sigma_1^2 \,+\, \sigma_2^2\cdot\mbox{tr}_x^2\left(B_{12}\right) \,+\, \sigma_3^2}}\,,
\label{TPS_01_3}
\end{eqnarray}

\noindent
\begin{eqnarray}
\alpha_{\varsigma}(s_2) \;=\;
\mbox{sign}\left(r_{12}\right)\cdot
\frac{\sigma_2}{\sigma_1 \, \sigma_3} \cdot
\frac{\sigma_1^2\,r_{11}\,-\, \sigma_3^2 \,r_{22}}
{\sqrt{\sigma_1^2 \,+\, \sigma_2^2\cdot\mbox{tr}_x^2\left(B_{12}\right) \,+\, \sigma_3^2}}\,,
\label{TPS_01_4}
\end{eqnarray}

\noindent
\begin{eqnarray}
\beta_{\varsigma}(s_3) \;=\; 
\left| r_{12} \right| \cdot
\frac{\sigma_3}{\sigma_1 \, \sigma_2} \cdot
\frac{\sigma_1^2 \,+\, \sigma_2^2\cdot\mbox{tr}_x^2\left(B_{12}\right)}
{\sqrt{\sigma_1^2 \,+\, \sigma_2^2\cdot\mbox{tr}_x^2\left(B_{12}\right) \,+\, \sigma_3^2}}\,,
\label{TPS_01_5}
\end{eqnarray}

\noindent
\begin{eqnarray}
\alpha_{\varsigma}(s_3) \;=\;-
\mbox{sign}\left(r_{12}\right)\cdot
\frac{\sigma_3}{\sigma_1 \, \sigma_2} \cdot
\frac{\,\sigma_1^2 \,r_{22}\,+\,
\sigma_2^2\cdot\mbox{tr}_x\left(B_{12}\right)\cdot\left(r_{22}\cdot\mbox{tr}_x\left(B_{12}\right) \,-\,1\right) }
{\sqrt{\sigma_1^2 \,+\, \sigma_2^2\cdot\mbox{tr}_x^2\left(B_{12}\right) \,+\, \sigma_3^2}}\,.
\label{TPS_01_6}
\end{eqnarray}

Let us assume in the following that BPM resolutions follow symmetry of the system,
which, in the periodic case, naturally mean that $\,\sigma_1 =\sigma_2 =\sigma_3$. 
In this situation $\,\beta_{\varsigma}(s_1)\,$ and $\,\beta_{\varsigma}(s_3)\,$
are always equal to each other and, it seems, it is the only symmetry which does not
require additional assumptions about coefficients of the cell transport matrix $\,B_{12}$.

The error betatron functions will be cell periodic (will stay unchanged after transport
through the first half of the system), if and only if

\noindent
\begin{eqnarray}
\mbox{tr}_x^2\left(B_{12}\right) \;=\; 1\,,
\label{TPS_010_1}
\end{eqnarray}

\noindent
and, if (\ref{TPS_010_1}) is satisfied, then

\noindent
\begin{eqnarray}
\cos(\mu_p) \;=\; \pm \frac{1}{2}
\label{TPS_010_2}
\end{eqnarray}

\noindent
and, therefore, 

\noindent
\begin{eqnarray}
\sin(3 \mu_p) \,=\, \sin(\mu_p) \cdot \left( 4 \cos^2(\mu_p)\,-\,1\right) \,=\, 0\,,
\label{TPS_010_3}
\end{eqnarray}

\noindent
where $\,\mu_p\,$ is the cell phase advance corresponding to the periodic betatron
function.

The error betatron functions will be two cell periodic (will stay unchanged after transport
through the whole system), if and only if

\noindent
\begin{eqnarray}
\mbox{tr}_x^3\left(B_{12}\right) \;=\; \mbox{tr}_x\left(B_{12}\right)\,,
\label{TPS_010_4}
\end{eqnarray}

\noindent
which, when compared with (\ref{TPS_010_1}), gives equation

\noindent
\begin{eqnarray}
\mbox{tr}_x\left(B_{12}\right) \;=\; 0\,,
\label{TPS_010_5}
\end{eqnarray}

\noindent
as the condition for the ``true two cell periodicity''
(two cell periodic, but not one cell periodic).
This condition means that the transverse part of the total system matrix $\,B_{12}^2\,$
(two by two submatrix located in the left upper corner) is equal
to the minus identity matrix for which arbitrary incoming
beta and alpha functions will be transported without changes through the system,
but the error betatron functions will also bring 
the sum of the beta function at the BPM locations to the minimal possible value.  

To finish the discussion about error betatron functions let us note, that if in the matrix $\,B_{12}$ 
the first two diagonal coefficients are to equal each other ($\,r_{11} \,=\, r_{22}\,$), then

\noindent
\begin{eqnarray}
\alpha_{\varsigma}(s_3) = -\alpha_{\varsigma}(s_1),
\hspace{1.0cm}
\alpha_{\varsigma}(s_2) = 0,
\label{TPS_010_7}
\end{eqnarray}

\noindent
and one may say that in this situation the error betatron function becomes mirror symmetric.

For the error energy spread and the error coordinate and momentum
dispersions we have in the case of the two periodic measurement system
the following formulas

\noindent
\begin{eqnarray}
\Delta^2_{\varsigma} \;=\;
\frac{\sigma_1^2 \,+\, \sigma_2^2\cdot\mbox{tr}_x^2\left(B_{12}\right) \,+\, \sigma_3^2}
{\left(r_{16}\,-\,r_{52}\right)^2}\,,
\label{TPS_02}
\end{eqnarray}

\noindent
\begin{eqnarray}
\eta_{x, \varsigma}(s_1) \;=\;
\frac{ \sigma_1^2 \, \left(r_{16}\,-\,r_{52}\right)}
{\sigma_1^2 \,+\, \sigma_2^2\cdot\mbox{tr}_x^2\left(B_{12}\right) \,+\, \sigma_3^2}\,,
\label{TPS_02_1}
\end{eqnarray}

\noindent
\begin{eqnarray}
\eta_{p, \varsigma}(s_1) \;=
\nonumber
\end{eqnarray}
\noindent
\begin{eqnarray}
= -
\frac{\sigma_1^2 \left( r_{16} + r_{11} \left(r_{16} - r_{52} \right)\right) +
\sigma_2^2 \cdot\mbox{tr}_x\left(B_{12}\right)\cdot
\left(\left(\mbox{tr}_x\left(B_{12}\right) + 1\right) r_{16} - r_{52} \right)
+ \sigma_3^2  r_{16}}
{r_{12} \cdot \left(\sigma_1^2 \,+\, \sigma_2^2\cdot\mbox{tr}_x^2\left(B_{12}\right) \,+\, \sigma_3^2\right)}\,,
\label{TPS_02_2}
\end{eqnarray}

\noindent
\begin{eqnarray}
\eta_{x, \varsigma}(s_2) \;=\;-
\frac{ \sigma_2^2 \cdot\mbox{tr}_x\left(B_{12}\right) \cdot\left(r_{16}\,-\,r_{52}\right)}
{\sigma_1^2 \,+\, \sigma_2^2\cdot\mbox{tr}_x^2\left(B_{12}\right) \,+\, \sigma_3^2}\,,
\label{TPS_02_3}
\end{eqnarray}

\noindent
\begin{eqnarray}
\eta_{p, \varsigma}(s_2) \;=
\nonumber
\end{eqnarray}
\noindent
\begin{eqnarray}
= -
\frac{
\sigma_1^2 r_{16} +
\sigma_2^2 \cdot \mbox{tr}_x\left(B_{12}\right) \cdot
\left( 
\mbox{tr}_x\left(B_{12}\right) \cdot r_{52} + 
r_{22} \left(r_{16}-r_{52}\right)\right)
+ \sigma_3^2  r_{52}}
{r_{12} \cdot \left(\sigma_1^2 \,+\, \sigma_2^2\cdot\mbox{tr}_x^2\left(B_{12}\right) \,+\, \sigma_3^2\right)}\,,
\label{TPS_02_4}
\end{eqnarray}

\noindent
\begin{eqnarray}
\eta_{x, \varsigma}(s_3) \;=\;
\frac{ \sigma_3^2 \, \left(r_{16}\,-\,r_{52}\right)}
{\sigma_1^2 \,+\, \sigma_2^2\cdot\mbox{tr}_x^2\left(B_{12}\right) \,+\, \sigma_3^2}\,,
\label{TPS_02_5}
\end{eqnarray}

\noindent
\begin{eqnarray}
\eta_{p, \varsigma}(s_3) \;=
\nonumber
\end{eqnarray}
\noindent
\begin{eqnarray}
= -
\frac{\sigma_1^2  r_{52}+
\sigma_2^2 \cdot\mbox{tr}_x\left(B_{12}\right)\cdot
\left(\left(\mbox{tr}_x\left(B_{12}\right) + 1\right) r_{52} - r_{16} \right)
+\sigma_3^2 \left( r_{52} + r_{22} \left(r_{52} - r_{16} \right)\right) 
}
{r_{12} \cdot \left(\sigma_1^2 
\,+\, \sigma_2^2\cdot\mbox{tr}_x^2\left(B_{12}\right) \,+\, \sigma_3^2\right)}\,,
\label{TPS_02_6}
\end{eqnarray}

\noindent
and, if resolutions of all three BPMs will be equal, the error dispersion
will satisfy the equality

\noindent
\begin{eqnarray}
\eta_{x, \varsigma}(s_1) \;=\;
\eta_{x, \varsigma}(s_3)\,.
\label{TPS_02_7}
\end{eqnarray}

The condition for the error dispersion to be cell periodic
is more restrictive than for the error betatron functions, namely

\noindent
\begin{eqnarray}
\mbox{tr}_x\left(B_{12}\right) \;=\; -1\,,
\label{TPS_02_8}
\end{eqnarray}

\noindent
and the condition for the ``true two cell periodicity'' is

\noindent
\begin{eqnarray}
\mbox{tr}_x\left(B_{12}\right) \;=\; -2\,,
\label{TPS_02_9}
\end{eqnarray}

\noindent
which does not lead to any noticeable symmetry of the error betatron functions and
which means that the transverse part of the cell matrix $\,B_{12}\,$ is equal to the
sum of the minus identity matrix plus some nilpotent matrix $\,N\,$
($\,N^2 = 0\,$).
 
Note that under condition (\ref{TPS_02_8}) we have for the periodic cell
phase advance the following relations

\noindent
\begin{eqnarray}
\cos(\mu_p) \,=\, -\frac{1}{2} ,
\hspace{1.0cm}
\sin(3 \mu_p / 2) \,=\, \sin(\mu_p / 2) \cdot \left( 2 \cos(\mu_p)\,+\,1\right) \,=\, 0.
\label{TPS_02_10}
\end{eqnarray}

\subsection{Cell Followed by Switched Cell}

\hspace{0.5cm}
If a magnet system between the second and the third BPMs repeats the magnet
system between the first and the second BPMs but with switched
directions of dipole magnets, then 

\noindent
\begin{eqnarray}
B_{23}\;=\;
\left(
\begin{array}{ccc}
r_{11} & r_{12}& -r_{16}\\
r_{21} & r_{22}& -r_{26}\\
0 & 0 & 1
\end{array}
\right) .
\label{SWT_01}
\end{eqnarray}

\noindent
In analogy with transition from mirror symmetric to mirror antisymmetric case,
the transverse error emittance and the error betatron functions
remain the same as for the two periodic measurement system, but,
in contrast to mirror antisymmetric case, this system still can be used
for the energy jitter measurement if

\noindent
\begin{eqnarray}
r_{16}\,+\,r_{52} \;\neq\;0 \,,
\label{SWT_01_01}
\end{eqnarray}
 
\noindent
which, in particular, forbids the magnet system between
the first and the second BPMs to be mirror symmetric by itself.

For this measurement system we obtain

\noindent
\begin{eqnarray}
\Delta^2_{\varsigma} \;=\;
\frac{\sigma_1^2 \,+\, \sigma_2^2\cdot\mbox{tr}_x^2\left(B_{12}\right) \,+\, \sigma_3^2}
{\left(r_{16}\,+\,r_{52}\right)^2}\,,
\label{SWT_02}
\end{eqnarray}

\noindent
\begin{eqnarray}
\eta_{x, \varsigma}(s_1) \;=\; -
\frac{ \sigma_1^2 \, \left(r_{16}\,+\,r_{52}\right)}
{\sigma_1^2 \,+\, \sigma_2^2\cdot\mbox{tr}_x^2\left(B_{12}\right) \,+\, \sigma_3^2}\,,
\label{SWT_03}
\end{eqnarray}

\noindent
\begin{eqnarray}
\eta_{p, \varsigma}(s_1) \;=
\nonumber
\end{eqnarray}
\noindent
\begin{eqnarray}
= -
\frac{\sigma_1^2 \left( r_{16} - r_{11} \left(r_{16} + r_{52} \right)\right) +
\sigma_2^2 \cdot\mbox{tr}_x\left(B_{12}\right)\cdot
\left(\left(\mbox{tr}_x\left(B_{12}\right) - 1\right) r_{16} - r_{52} \right)
+ \sigma_3^2  r_{16}}
{r_{12} \cdot \left(\sigma_1^2 \,+\, \sigma_2^2\cdot\mbox{tr}_x^2\left(B_{12}\right) \,+\, \sigma_3^2\right)}\,,
\label{SWT_04}
\end{eqnarray}

\noindent
\begin{eqnarray}
\eta_{x, \varsigma}(s_2) \;=\;
\frac{ \sigma_2^2 \cdot\mbox{tr}_x\left(B_{12}\right) \cdot\left(r_{16}\,+\,r_{52}\right)}
{\sigma_1^2 \,+\, \sigma_2^2\cdot\mbox{tr}_x^2\left(B_{12}\right) \,+\, \sigma_3^2}\,,
\label{SWT_05}
\end{eqnarray}

\noindent
\begin{eqnarray}
\eta_{p, \varsigma}(s_2) \;=
\nonumber
\end{eqnarray}
\noindent
\begin{eqnarray}
= 
\frac{
\sigma_1^2 r_{16} -
\sigma_2^2 \cdot \mbox{tr}_x\left(B_{12}\right) \cdot
\left( 
\mbox{tr}_x\left(B_{12}\right) \cdot r_{52} - 
r_{22} \left(r_{16}+r_{52}\right)\right)
- \sigma_3^2  r_{52}}
{r_{12} \cdot \left(\sigma_1^2 \,+\, \sigma_2^2\cdot\mbox{tr}_x^2\left(B_{12}\right) \,+\, \sigma_3^2\right)}\,,
\label{SWT_06}
\end{eqnarray}

\noindent
\begin{eqnarray}
\eta_{x, \varsigma}(s_3) \;=\; -
\frac{ \sigma_3^2 \, \left(r_{16}\,+\,r_{52}\right)}
{\sigma_1^2 \,+\, \sigma_2^2\cdot\mbox{tr}_x^2\left(B_{12}\right) \,+\, \sigma_3^2}\,,
\label{SWT_07}
\end{eqnarray}

\noindent
\begin{eqnarray}
\eta_{p, \varsigma}(s_3) \;=
\nonumber
\end{eqnarray}
\noindent
\begin{eqnarray}
= 
\frac{\sigma_1^2  r_{52}+
\sigma_2^2 \cdot\mbox{tr}_x\left(B_{12}\right)\cdot
\left(\left(\mbox{tr}_x\left(B_{12}\right) - 1\right) r_{52} - r_{16} \right)
+\sigma_3^2 \left( r_{52} - r_{22} \left(r_{52} + r_{16} \right)\right) 
}
{r_{12} \cdot \left(\sigma_1^2 
\,+\, \sigma_2^2\cdot\mbox{tr}_x^2\left(B_{12}\right) \,+\, \sigma_3^2\right)}\,,
\label{SWT_08}
\end{eqnarray}

\noindent
and one sees that for equal BPM resolutions the property

\noindent
\begin{eqnarray}
\eta_{x, \varsigma}(s_1) \;=\; 
\eta_{x, \varsigma}(s_3)
\label{SWT_09}
\end{eqnarray}

\noindent
is still preserved.

There is no reasons to expect that coordinate error dispersion and
simultaneously momentum error dispersion could stay constant at all three BPM locations
(analogy of cell periodicity for the two cell measurement system) and, as one
can check, there is no solution for that. Nevertheless, both error dispersions still
can stay unchanged after transport through the whole system, if

\noindent
\begin{eqnarray}
\mbox{tr}_x\left(B_{12}\right) \,=\, 1
\hspace{0.7cm} 
\mbox{or}
\hspace{0.7cm} 
\mbox{tr}_x\left(B_{12}\right) \,=\, 2.
\label{SWT_10}
\end{eqnarray}

\section{Periodic Measurement Systems}

\hspace{0.5cm}
Let us consider a measurement system constructed from
$n$ identical cells assuming that the
cell transfer matrix allows periodic beam transport
with phase advance per cell $\,\mu_p\,$ being not a multiple of
$\,180^{\circ}$. Additionally, we will assume that BPMs placed in our
beam line deliver uncorrelated readings, all with the same rms
resolution $\,\sigma_{bpm}$. 

Let us first consider the case when we have
one BPM per cell (identically positioned in all cells) 
with the periodic betatron function and the periodic dispersion function
at the BPM locations
equal to $\,\beta_p(s_1)\,$ and $\,\eta_{x,p}(s_1) \neq 0\,$ respectively. 

In this situation the formulas for the error transverse emittance and
the error betatron function remain the same as was already published in \cite{ThePaper}, 
and the square of the 
error energy spread is given by the following
expression

\noindent
\begin{eqnarray}
\Delta^2_{\varsigma} \;=\;
\frac{\sigma^2_{bpm}}{n \, \eta_{x,p}^2(s_1)}
\cdot \varrho_n(\mu_p) \,,
\label{disp_0}
\end{eqnarray}

\noindent
where the function

\noindent
\begin{eqnarray}
\varrho_n(\mu_p) \;=\;
\frac{1 \;+\; \dfrac{1}{n}\cdot \dfrac{\sin(n \mu_p)}{\sin(\mu_p)}}
{\;
1 \;+\; \dfrac{1}{n} \cdot\dfrac{\sin(n \mu_p)}{\sin(\mu_p)} \;-\; 
2 \left(\dfrac{1}{n} \cdot \dfrac{\sin(n \mu_p / 2)}{\sin(\mu_p / 2)}\right)^2
\;}
\label{disp_1}
\end{eqnarray}

\noindent
is defined only for $\,n \geq 3$.\footnote{For 
$\,n = 1, 2\,$ the denominator in the formula (\ref{disp_1}) is equal
to zero independent of the value of the cell phase advance $\mu_p$}
Note that for $\,n \geq 3\,$ this function could
be extended by continuity for all $\,\mu_p\,$ not multiple of $360^{\circ}$
where it becomes unbounded.\footnote{The nonnegative denominator in the formula (\ref{disp_1}) is 
equal to zero
not only when $\,\mu_p\,$ is a multiple of $360^{\circ}$, but also when
$n$ is even and, simultaneously, $\,\mu_p\,$ an odd multiple of $180^{\circ}$.}

The coordinate and momentum error dispersions $\,\eta_{x, \varsigma}\,$ and 
$\,\eta_{p, \varsigma}\,$ at the BPM locations
are given below

\noindent
\begin{eqnarray}
\eta_{x, \varsigma}(s_k) \;=\; \eta_{x,p}(s_1)
\cdot
\left(
1 \;-\; \omega_n(\mu_p)
\cdot
\cos\left(\tfrac{n + 1 - 2 k}{2} \, \mu_p\right)
\right)\,,
\label{disp_2}
\end{eqnarray}

\noindent
\begin{eqnarray}
\eta_{p, \varsigma}(s_k) \;=\; \eta_{p,p}(s_1) \;-\; \eta_{x,p}(s_1)
\cdot \frac{\omega_n(\mu_p)}{\beta_p(s_1)}
\left(
\sin\left(\tfrac{n + 1 - 2 k}{2} \, \mu_p\right) \;-
\right.
\nonumber
\end{eqnarray}
\begin{eqnarray}
\left.
-\;
\alpha_p(s_1)
\cdot
\cos\left(\tfrac{n + 1 - 2 k}{2} \, \mu_p\right)
\right)\,,
\label{disp_3}
\end{eqnarray}

\noindent
\begin{eqnarray}
\omega_n(\mu_p) \;=\;
2 \, \left(\dfrac{1}{n}\cdot \dfrac{\sin(n \mu_p / 2)}{\sin(\mu_p /2)}\right)
\left(1 \;+\; \dfrac{1}{n} \cdot\dfrac{\sin(n \mu_p)}{\sin(\mu_p)}\right)^{-1} \,,
\label{disp_4}
\end{eqnarray}

\noindent
and one sees that while the coordinate error dispersion 
$\,\eta_{x, \varsigma}\,$ always have mirror symmetry

\noindent
\begin{eqnarray}
\eta_{x, \varsigma} (s_k) \;=\;
\eta_{x, \varsigma} (s_{n + 1 - k})\,,
\hspace{1cm}
k \;=\; 1,\,\ldots,\,n ,
\label{disp_5}
\end{eqnarray}

\noindent
the momentum error dispersion will be mirror antisymmetric

\noindent
\begin{eqnarray}
\eta_{p, \varsigma} (s_k) \;=\;
-\eta_{p, \varsigma} (s_{n + 1 - k})\,,
\hspace{1cm}
k \;=\; 1,\,\ldots,\,n 
\label{disp_6}
\end{eqnarray}

\noindent
only in the case when $\,\alpha_p(s_1)\,=\,0\,$ and $\,\eta_{p, p}(s_1)\,=\,0$.

Note, that the mean value of the coordinate error dispersion
and the mean value of its squares satisfy the following relations 

\noindent
\begin{eqnarray}
\frac{1}{n} \,\sum_{k = 1}^{n} \eta_{x, \varsigma}  (s_k) \;=\;
\frac{\eta_{x, p} (s_1)}{\varrho_n(\mu_p)}\,,
\hspace{1cm}
\frac{1}{n} \,\sum_{k = 1}^{n} \eta_{x, \varsigma}^2  (s_k) \;=\;
\frac{\eta_{x, p}^2 (s_1)}{\varrho_n(\mu_p)}\,.
\label{disp_7}
\end{eqnarray}

\begin{figure} 
\centerline{\hbox{\psfig{figure=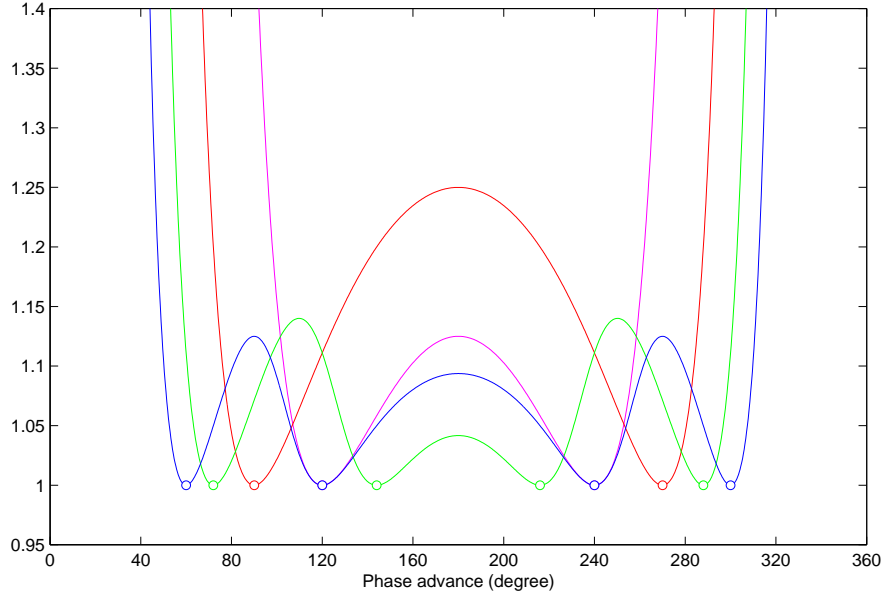,width=14.0cm}}}
\caption{Functions $\,\varrho_n(\mu_p)\,$ shown for
$\,n = 3, 4, 5, 6\,$ (magenta, red, green and blue curves respectively).}
\label{fig_P2_1}
\end{figure}

The function $\,\varrho_n(\mu_p)\,$ is never smaller than one and is equal to one
(reaches its minimum) only in the points 

\noindent
\begin{eqnarray}
\mu_p = k \, \frac{360^{\circ}}{n} \;(\mbox{mod} \;360^{\circ}) ,
\hspace{0.3cm}
k =
\left\{
\begin{array}{ll}
1, \ldots, n - 1 & \mbox{\;\,if $\,n\,$ is odd}\\
 1, \ldots, \frac{n}{2} - 1, \frac{n}{2} + 1, \ldots, n - 1 &\mbox{\;\,if $\,n\,$ is even}
\end{array}
\right.
\label{ddddd_1}
\end{eqnarray}

\noindent
in which error dispersion coincides with periodic dispersion and 
which seem to be good candidates to be selected for improving
resolution of the energy jitter measurement (see figure 3), if we are free in
choosing the cell phase advance while, for some reasons, the dispersion at the
BPM location has to stay unchanged. But, when we optimize a cell
in which periodic dispersion at the BPM location is by itself function of
the cell phase advance, the situation, of course, changes. Let us, like
in \cite{ThePaper}, consider a thin lens FODO cell of the length $L$
in which two identical thin lens dipoles with transfer matrix

\noindent
\begin{eqnarray}
\left(
\begin{array}{ccc}
1 & 0 & 0 \\
0 & 1 & \varphi /2\\
0 & 0 & 1
\end{array}
\right)
\label{ddddd_2}
\end{eqnarray}

\noindent
are inserted in the middle of drift spaces
separating the focusing and defocusing lenses.
Let us assume that the BPM is placed in the ``center" of
the focusing lens with the periodic dispersion at this locations being

\noindent
\begin{eqnarray}
\eta_{x, p}(s_1)\;=\;
\eta_{+}\;=\; \frac{L \varphi}{4} \cdot \frac{1 + \frac{1}{2} \sin(\mu_p / 2)}{\sin^2(\mu_p / 2)}\,,
\label{ddddd_3}
\end{eqnarray}

\noindent
where $\,\varphi\,$ is the cell deflection angle.

In this situation we can write

\noindent
\begin{eqnarray}
\Delta^2_{\varsigma} \;=\;
\frac{\sigma^2_{bpm}}{n} \left( \frac{4}{L \varphi}\right)^2
\cdot \Psi_n(\mu_p) \,,
\label{ddddd_4}
\end{eqnarray}

\noindent
where functions $\,\Psi_n\,$ depend only on the cell phase advance $\,\mu_p\,$ 
and are converging (from above) to the function

\noindent
\begin{eqnarray}
\Psi_{\infty}(\mu_p) \,=\,
\frac{\sin^4(\mu_p / 2)}{\left(1 + \frac{1}{2} \sin(\mu_p / 2) \right)^2}
\label{ddddd_5}
\end{eqnarray}

\noindent
as $n$ goes to infinity. 

The functions $\,\Psi_n \left( \mu_p \right)\,$ 
for $\,n = 3, 4, 5, 6\,$
are plotted in figure 4 
together with their values in the points (\ref{ddddd_1})
shown as small circles at the corresponding curves.
One sees that, again like in \cite{ThePaper}, there is nothing really special about
points (\ref{ddddd_1}) except the trivial fact
that all of them belong to the graph of the function
$\,\Psi_{\infty}\,$. 

\begin{figure} 
\centerline{\hbox{\psfig{figure=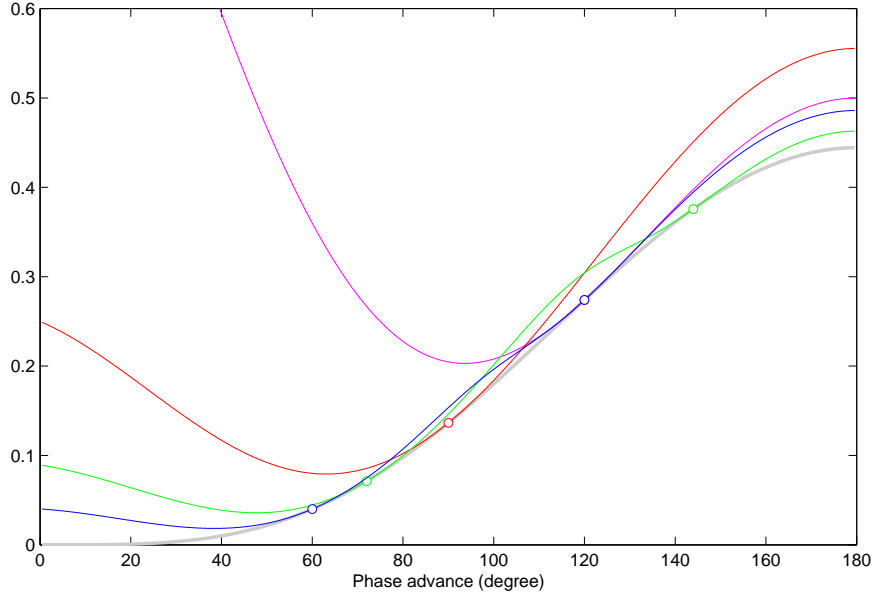,width=14.0cm}}}
\caption{Functions $\,\Psi_n(\mu_p)\,$ shown for
$\,n = 3, 4, 5, 6\,$ (magenta, red, green and blue curves respectively).
The gray curve shows function $\Psi_{\infty}(\mu_p)$.}
\label{fig_P2_2}
\end{figure}

\begin{figure} 
\centerline{\hbox{\psfig{figure=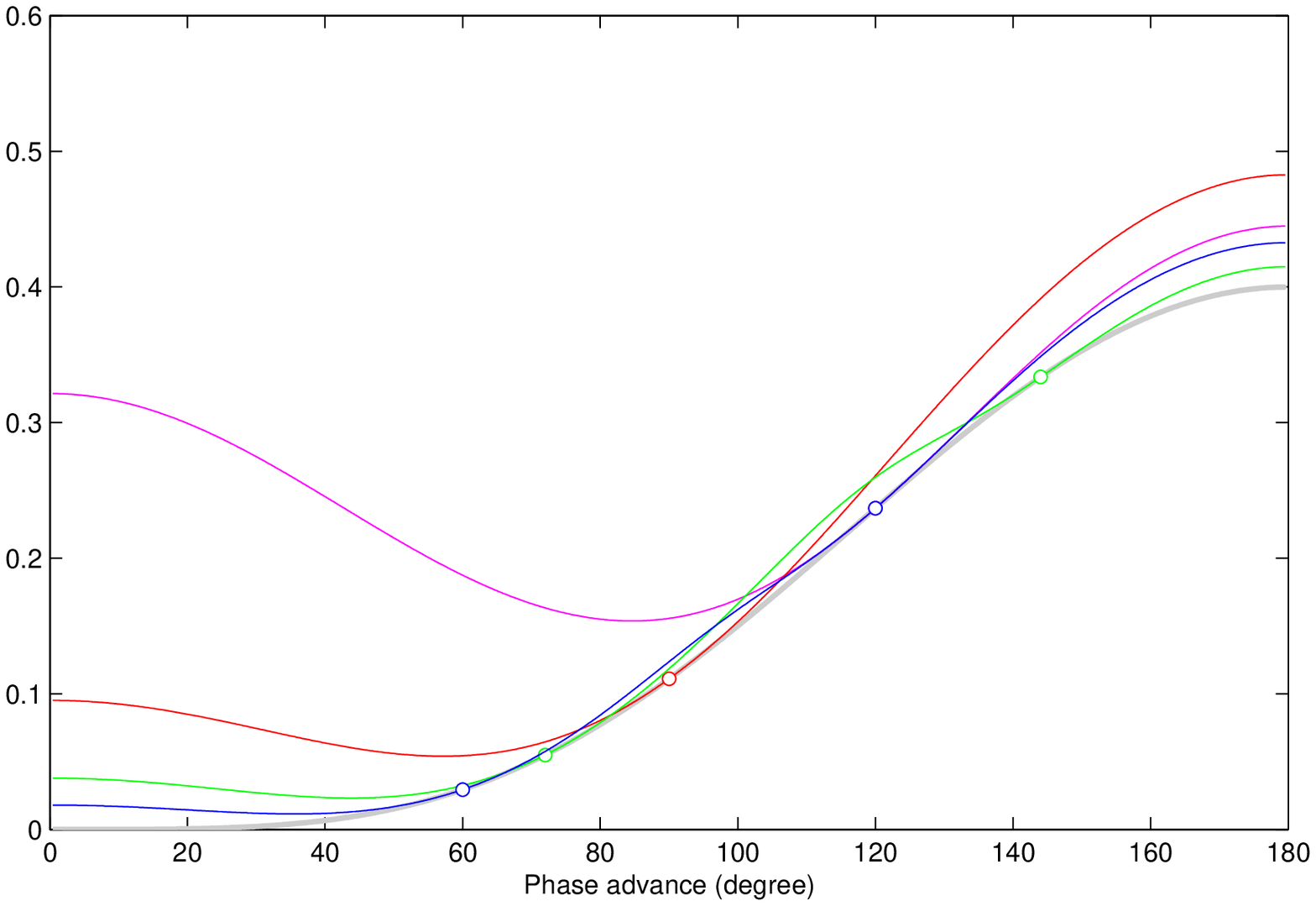,width=14.0cm}}}
\caption{Functions $\,\Phi_n(\mu_p)\,$ shown for
$\,n = 3, 4, 5, 6\,$ (magenta, red, green and blue curves respectively).
The gray curve shows function $\Phi_{\infty}(\mu_p)$.}
\label{fig_P2_3}
\end{figure}

Before switching to the situation when we have two BPMs per cell let us
rewrite expression (\ref{disp_1}) for the function $\,\varrho_n(\mu_p)\,$ in the form

\noindent
\begin{eqnarray}
\varrho_n(\mu_p) \,=\,
\left(
1 \,-\, \frac{\beta_p(s_1)}{2 \,m_p\left(\beta_{\varsigma}, \, \beta_p \right) \,\eta_{x,p}^2(s_1)}
\cdot
I_x\left(\beta_{\varsigma}, \,\eta_{x,\varsigma} - \eta_{x,p}, \,\eta_{p,\varsigma} - \eta_{p,p}\right)
\right)^{-1} ,
\label{disp_8}
\end{eqnarray}

\noindent
where

\noindent
\begin{eqnarray}
m_p\left(\beta_{\varsigma}, \, \beta_p \right) =
\left(1 \,-\, 
\left(\frac{1}{n}\cdot \frac{\sin(n \mu_p)}{\sin(\mu_p)}\right)^2
\right)^{-\frac{1}{2}}
\label{disp_8_001}
\end{eqnarray}

\noindent
is the mismatch between the error and the periodic betatron functions
(even so we do not assume, in general, periodic betatron functions and/or periodic
dispersion being the design
betatron functions and/or design dispersion matched to our beam line) and

\noindent
\begin{eqnarray}
I_x\left(\beta_{\varsigma}, \,\eta_{x,\varsigma} - \eta_{x,p}, \,\eta_{p,\varsigma} - \eta_{p,p}\right)
\;=
\nonumber
\end{eqnarray}

\noindent
\begin{eqnarray}
=\;
\gamma_{\varsigma} \left(\eta_{x,\varsigma} - \eta_{x,p}\right)^2
\,+\, 2 \alpha_{\varsigma} \left(\eta_{x,\varsigma} - \eta_{x,p}\right)
\left(\eta_{p,\varsigma} - \eta_{p,p}\right)
\,+\,\beta_{\varsigma} \left(\eta_{p,\varsigma} - \eta_{p,p}\right)^2 \;=
\nonumber
\end{eqnarray}

\noindent
\begin{eqnarray}
= \;
\frac{4\, \eta_{x,p}^2(s_1)\, m_p^3\left(\beta_{\varsigma}, \, \beta_p \right)}{\beta_p(s_1)}
\cdot\left(
\frac{1}{n}\cdot \frac{\sin(n \mu_p / 2)}{\sin(\mu_p /2)}\right)^2
\cdot\left(1 \;-\; \frac{1}{n} \cdot\frac{\sin(n \mu_p)}{\sin(\mu_p)}\right)
\label{disp_9}
\end{eqnarray}

\noindent
is the difference 
between periodic and error dispersions measured by using
the Courant-Snyder invariant formed out of error Twiss parameters.

Note, for completeness, that if one will express the difference 
between periodic and error dispersions using
Courant-Snyder invariant formed using periodic Twiss parameters,
then one will have the following relation 

\noindent
\begin{eqnarray}
I_x\left(\beta_p, \,\eta_{x,\varsigma} - \eta_{x,p}, \,\eta_{p,\varsigma} - \eta_{p,p}\right)
\;=
\nonumber
\end{eqnarray}

\noindent
\begin{eqnarray}
=\;
m_p\left(\beta_{\varsigma}, \, \beta_p \right)
\left(1 \;-\; \frac{1}{n} \cdot\frac{\sin(n \mu_p)}{\sin(\mu_p)}\right)
\cdot
I_x\left(\beta_{\varsigma}, \,\eta_{x,\varsigma} - \eta_{x,p}, \,\eta_{p,\varsigma} - \eta_{p,p}\right)
\;=
\nonumber
\end{eqnarray}

\noindent
\begin{eqnarray}
=\;
\sqrt{
\dfrac{1 \;-\; \dfrac{1}{n} \cdot\dfrac{\sin(n \mu_p)}{\sin(\mu_p)}}
{1 \;+\; \dfrac{1}{n} \cdot\dfrac{\sin(n \mu_p)}{\sin(\mu_p)}}
}
\cdot
I_x\left(\beta_{\varsigma}, \,\eta_{x,\varsigma} - \eta_{x,p}, \,\eta_{p,\varsigma} - \eta_{p,p}\right)\,.
\label{disp_8_002}
\end{eqnarray}

Let us now turn to the situation when we have two BPMs per cell with 
$\,\theta\,$ being the phase shift between the first and second BPM
location. 

In this situation the square of the error energy spread can be expressed as 

\noindent
\begin{eqnarray}
\Delta^2_{\varsigma} =
\frac{\sigma^2_{bpm}}{n  \left(\eta_{x,p}^2(s_1) + \eta_{x,p}^2(s_2)\right)}
\cdot 
\varpi_n\,,
\label{disp_2_0}
\end{eqnarray}

\noindent
where multiplier

\noindent
\begin{eqnarray}
\varpi_n
=
\left(
1 - \frac{\beta_p(s_1)+\beta_p(s_2)}{2 m_p\left(\beta_{\varsigma},  \beta_p \right) 
\left(\eta_{x,p}^2(s_1)+\eta_{x,p}^2(s_2)\right)}
\cdot
I_x\left(\beta_{\varsigma}, \eta_{x,\varsigma} - \eta_{x,p}, \eta_{p,\varsigma} - \eta_{p,p}\right)
\right)^{-1} 
\label{disp_2_0_1}
\end{eqnarray}

\noindent
has a form which is very similar to (\ref{disp_8}) with 

\noindent
\begin{eqnarray}
I_x\left(\beta_{\varsigma}, \,\eta_{x,\varsigma} - \eta_{x,p}, \,\eta_{p,\varsigma} - \eta_{p,p}\right)
\;=
\nonumber
\end{eqnarray}

\noindent
\begin{eqnarray}
=\;
\gamma_{\varsigma} \left(\eta_{x,\varsigma} - \eta_{x,p}\right)^2
\,+\, 2 \alpha_{\varsigma} \left(\eta_{x,\varsigma} - \eta_{x,p}\right)
\left(\eta_{p,\varsigma} - \eta_{p,p}\right)
\,+\,\beta_{\varsigma} \left(\eta_{p,\varsigma} - \eta_{p,p}\right)^2 \;=
\nonumber
\end{eqnarray}

\noindent
\begin{eqnarray}
= 
\frac{
4 
\left(
\eta_{x,p}^2(s_1) + \eta_{x,p}^2(s_2)
\right) 
m_p^3
\left(
\beta_{\varsigma},  \beta_p 
\right)
}
{
\beta_p(s_1)+ \beta_p(s_2)
}
\cdot
\left(
\frac{1}{n} 
\cdot 
\frac{\sin(n \mu_p / 2)}
{\sin(\mu_p /2)}
\right)^2
\Bigg[
\left(
1 - \frac{1}{n} \cdot 
\frac{\sin(n \mu_p)}{\sin(\mu_p)}
\right)
\cdot
\Bigg.
\nonumber
\end{eqnarray}

\noindent
\begin{eqnarray}
\cdot
\frac{
\beta_p(s_1)\, \eta_{x,p}^2(s_1) 
\, + \,  
2 \cos(\theta) \sqrt{\beta_p(s_1)\,\beta_p(s_2)} \,\eta_{x,p}(s_1)\,\eta_{x,p}(s_2)
\, + \,
\beta_p(s_2)\, \eta_{x,p}^2(s_2)
}
{
\left(
\beta_p(s_1)\,+\,\beta_p(s_2) 
\right)
\left(\eta_{x,p}^2(s_1) \,+\,\eta_{x,p}^2(s_2)
\right)
}
\;+
\nonumber
\end{eqnarray}

\noindent
\begin{eqnarray}
\Bigg.
+\;
2 \sin^2(\theta)
\frac{\beta_p(s_1)\,\beta_p(s_2)}
{\left( \beta_p(s_1) + \beta_p(s_2) \right)^2}
\cdot
\left(
\frac{1}{n}\cdot \frac{\sin(n \mu_p)}{\sin(\mu_p)}
\right)
\Bigg]
\label{disp_2_9}
\end{eqnarray}

\noindent
and with the mismatch between the error and the periodic betatron functions
having now the following form

\noindent
\begin{eqnarray}
m_p\left(\beta_{\varsigma}, \, \beta_p \right) = \left(1 -
\left(1 - 4 \sin^2(\theta)
\frac{\beta_p(s_1)\beta_p(s_2)}{\left(\beta_p(s_1) + \beta_p(s_2)\right)^2}
\right)
\left(\frac{1}{n}\cdot \frac{\sin(n \mu_p)}{\sin(\mu_p)}\right)^2
\right)^{-\frac{1}{2}}\,.
\label{disp_2_1}
\end{eqnarray}

For a thin lens FODO cell with BPMs placed in the ``centers" of
focusing and defocusing lenses we have $\,\theta \,=\, \mu_p / 2\,$ and
the periodic beta function and the periodic dispersion at the BPM locations 
are equal to

\noindent
\begin{eqnarray}
\beta_{\pm}\;=\; L \cdot \frac{1 \,\pm\, \sin(\mu_p / 2)}{\sin(\mu_p)}\,,
\hspace{0.5cm}
\eta_{\pm}\;=\; 
\frac{L \varphi}{4} \cdot \frac{1 \,\pm\, \frac{1}{2} \sin(\mu_p / 2)}{\sin^2(\mu_p / 2)}\,.
\label{kkkkkkk_1}
\end{eqnarray}

\noindent
With these assumptions we can write 

\noindent
\begin{eqnarray}
\Delta^2_{\varsigma} \;=\;
\frac{\sigma^2_{bpm}}{n} \left( \frac{4}{L \varphi}\right)^2
\cdot \Phi_n(\mu_p) \,,
\label{kkkkkk_2}
\end{eqnarray}

\noindent
with functions $\Phi_n(\mu_p)$ converging to the function

\noindent
\begin{eqnarray}
\Phi_{\infty}(\mu_p) \,=\,
\frac{\sin^4(\mu_p / 2)}{2 + \frac{1}{2} \sin^2(\mu_p / 2)}
\label{kkkkkk_3}
\end{eqnarray}

\noindent
as $n$ goes to infinity. 

The functions $\,\Phi_n \left( \mu_p \right)\,$ 
for $\,n = 3, 4, 5, 6\,$
are plotted in figure 5 and one can see that though we are using 
two times larger number of BPMs, the energy resolution improves 
mainly in the region of the low phase advances, while
for the high phase advances it stays almost unchanged.
To understand the situation better, it is useful to look
at the figure 6 where the ratio of the limiting functions
$\,\Psi_{\infty}\,$ and $\,\Phi_{\infty}\,$
is shown.

\begin{figure} 
\centerline{\hbox{\psfig{figure=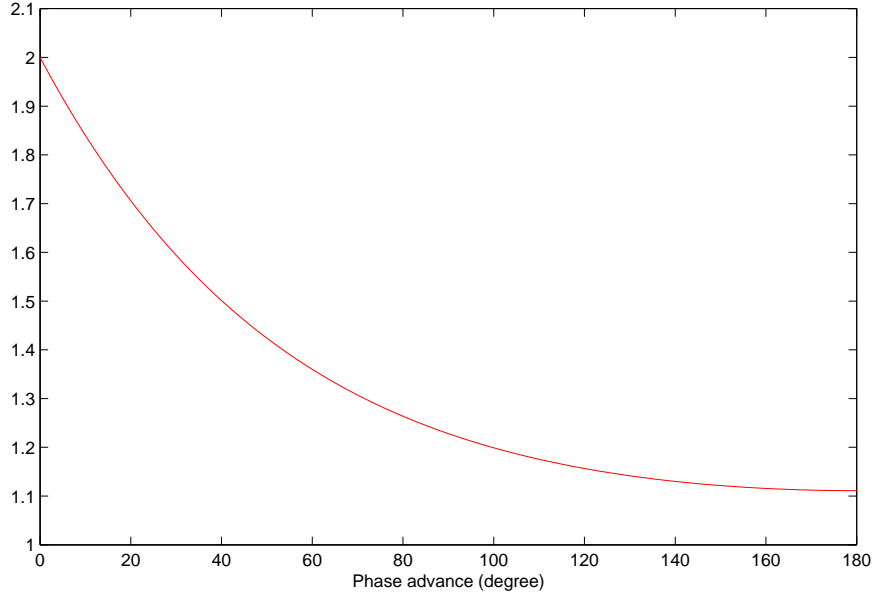,width=14.0cm}}}
\caption{Ratio $\,\Psi_{\infty}\, / \,\Phi_{\infty}$ as a function
of the phase advance $\,\mu_p$.}
\label{fig_P2_5}
\end{figure}

\section{Courant-Snyder Invariant as Error Estimator}

\hspace{0.5cm}
When we consider the jitter problem, the subject of our real interest 
is the actual difference $\,\delta \vec{z}_0\,$ between parameters of 
the instantaneous and the golden trajectory.
Our measurement system delivers 
us an estimate $\,\delta \vec{z}_{\varsigma}\,$ of this parameter, which 
includes the effect of the BPM reading errors $\,\vec{\,\varsigma}$. 

Thus, in the framework of the model considered,
the only information which we could obtain about the true difference 
$\,\delta \vec{z}_0\,$ is the statistical information based on
the properties of the random variable 
$\,\delta \vec{z}_{\varsigma} - \delta \vec{z}_0$,
which, due to our assumptions, has zero mean and whose statistical distribution 
does not depend on the actual value of $\,\delta \vec{z}_0\,$. 

It seems to be natural to use the module
$\,\left|\delta \varepsilon_{\varsigma} - \delta \varepsilon_0\right|\,$
as a numerical measure of the difference between estimated and true beam 
energies, but the quantitative measure 
of the difference $\,\delta \vec{z}_{\varsigma} - \delta \vec{z}_0\,$
from zero in the transverse phase space could be chosen differently.
One may simply use the Euclidean vector norm, but, as it was already stated 
in \cite{ThePaper}, the usage of the Courant-Snyder quadratic form has certain advantages.
For example, when one considers errors only in the reconstruction of 
the transverse orbit parameters in the beam line without dispersion,
the Courant-Snyder quadratic form is an invariant and
therefore all estimates based on it do not depend on the position of
the reconstruction point. And, as one will see below, even for dispersive 
particle motion the Courant-Snyder quadratic form is a ``much better
conserved quantity'' than the Euclidean norm.

\subsection{Transverse Jitter}

\hspace{0.5cm}
Let us first return to the situation whose study was already started 
in paper \cite{ThePaper}, where we considered errors in the reconstruction 
of transverse orbit parameters in the beam line without dispersion.

Let 
$\,\beta_0(r), \, \alpha_0(r)\,$ and $\,\gamma_0(r)\,$ be the 
design betatron functions, and

\noindent
\begin{eqnarray}
I_x (\,r, \, x, \, p\,)\; = \;
\gamma_0(r)\, x^{2}  \;+\; 2\, \alpha_0(r) \, x\, p 
\;+\; \beta_0(r) \, p^{2} 
\label{IFB_1}
\end{eqnarray}

\noindent
be the corresponding Courant-Snyder quadratic form. 

According to the above discussion, the object of our current interest is
the random variable

\noindent
\begin{eqnarray}
I_x^{\varsigma} \;=\; I_x ( \,r, \,\, \delta x_{\varsigma} -  \delta x_{0} \, ,
\, \delta p_{\varsigma}  -  \delta p_{0}  \, ) . 
\label{SEC_GD_1}
\end{eqnarray}

\noindent
The mean value of this random variable was already calculated in \cite{ThePaper}
and is equal

\noindent
\begin{eqnarray}
\big< \, I_x^{\varsigma} \, \big> 
\;=\; 2 \,\epsilon_{\varsigma} \, m_p(\beta_{\varsigma},\,\beta_0) \,,
\label{App1}
\end{eqnarray}

\noindent
where $\,m_p(\beta_{\varsigma},\,\beta_0)\,$ is the mismatch
between the error and the design betatron functions.
That is, probably, all what one can obtain without making additional 
assumptions about distribution of BPM reading errors. 

In this subsection we will assume that the random vector $\,\vec{\,\varsigma}\,$ 
has a multivariate normal distribution and will find not only 
higher order moments of the random variable $\,I_x^{\varsigma}\,$, but also its
probability density.

Calculations made in \cite{ThePaper} show that we can represent the
variable $\,I_x^{\varsigma}\,$ in the form

\noindent
\begin{eqnarray}
I_x^{\varsigma}
\; = \; 
\vec{\,\eta}^{\top}
K^{\top}(r) K(r)
\vec{\,\eta}\,,
\label{SEC_WC_7}
\end{eqnarray}

\noindent
where

\noindent
\begin{eqnarray}
K \,=\, T V_z M^{\top} R_{\varsigma}^{-1},
\hspace{0.5cm}
T\,=\,\left(
\begin{array}{cc}
1 / \sqrt{\beta_0} & 0\\
\alpha_0 / \sqrt{\beta_0} & \sqrt{\beta_0}
\end{array}
\right),
\hspace{0.5cm}
M\,=\,\left(
\begin{array}{cc}
a_1    &  c_1   \\
\vdots & \vdots \\
a_n    &  c_n
\end{array}
\right),
\label{SEC_WC_6}
\end{eqnarray}

\noindent
and the components of the vector 
$\,\vec{\,\eta} = R_{\varsigma}^{-\top} \vec{\,\varsigma}\,$ are independent standard
normal variables. 
The matrix $K^{\top} K$ is $n$ by $n$ matrix, but, as it was also shown 
in \cite{ThePaper}, it has only two nonzero eigenvalues, namely

\noindent
\begin{eqnarray}
\mu_{\pm}\;=\;
\epsilon_{\varsigma} \left(
m_p\left(\beta_{\varsigma},\,\beta_0\right)
\,\pm\,
\sqrt{m_p^2\left(\,\beta_{\varsigma},\,\beta_0\right)\,-\,1\,}
\right) .
\label{SEC_WC_11}
\end{eqnarray}

If $\,\vec{e}_{\pm}\,$ are the unit orthogonal eigenvectors of the symmetric matrix
$\,K^{\top} K\,$ corresponding to its nonzero eigenvalues $\,\mu_{\pm}\,$, then
we can rewrite (\ref{SEC_WC_7}) in the form

\noindent
\begin{eqnarray}
I_x^{\varsigma} \;=\; \mu_{+} \, \xi_{+}^2 \;+\; \mu_{-} \, \xi_{-}^2\,,
\label{SEC_GD_2}
\end{eqnarray}

\noindent
where 
$\,\xi_{\pm} =\vec{e}_{\pm}^{\,\top} \vec{\,\eta}\,$
are two independent standard normal variables.

With representation (\ref{SEC_GD_2})
calculation of all probabilistic characteristics of the random variable
$\,I_x^{\varsigma}\,$ becomes rather straightforward. 
For example, the following formula gives its variance

\noindent
\begin{eqnarray}
{\cal V} \left(\, I_x^{\varsigma} \, \right) \;=\;
\big< \, \left(I_x^{\varsigma}\right)^2 \, \big> 
\,-\, \big< \, I_x^{\varsigma} \, \big>^2 \;=\;
4 \,\epsilon_{\varsigma}^2 \, \left(
2 \,m_p^2(\beta_{\varsigma},\,\beta_0) \,-\,1 \right)\,.
\label{App2}
\end{eqnarray}

\noindent
Moreover, it is not very complicated to calculate
the probability density of this random variable
using, for example, results published in \cite{GradSolomon}.
This density $\,p(t)\,$ is equal to zero for negative values of its argument,
and for $\,t \geq 0\,$

\noindent
\begin{eqnarray}
p(t) \;=\; \frac{1}{2 \epsilon_{\varsigma}} \,
I_0 \left(\sqrt{m_p^2(\beta_{\varsigma},\,\beta_0) - 1} \,\, 
\frac{t}{2 \epsilon_{\varsigma}}\right)
\exp \left(-m_p(\beta_{\varsigma},\,\beta_0) \,\frac{t}{2 \epsilon_{\varsigma}}\right)\,,
\label{probD}
\end{eqnarray}

\noindent
where $I_0$ is the modified Bessel function of zero order.

Note that for $\,m_p(\beta_{\varsigma},\,\beta_0) = 1\,$ the density (\ref{probD}) becomes
the density of chi-square distribution with two degrees of freedom
and in this case the distribution function $\,F(t)\,$ can be
calculated in the explicit form

\noindent
\begin{eqnarray}
F(t) \;=\;\Pr \left(\, I_x^{\varsigma} \,\leq \,t \, \right)
\;=\; \int \limits_0^t \,p(\tau) \,d \tau 
\;=\;1 \,-\,\exp\left(-\frac{t}{2 \epsilon_{\varsigma}}\right)
\,.
\label{probD2}
\end{eqnarray}

\subsection{Transverse and Energy Jitter}

\hspace{0.5cm}
When the beam energy is included in both, measurement and dynamics,
the transverse motion could be separated into two parts:
dispersive motion and pure betatron oscillations. One can write

\noindent
\begin{eqnarray}
\left\{
\begin{array}{ccc}
\delta x_0 &=&\left(\delta x_0  \,-\, \delta \varepsilon_0 \cdot \eta_{x, 0} \right)
\,+\, \delta \varepsilon_0 \cdot \eta_{x, 0}\\
\delta p_0 &=& \left(\delta p_0  \,-\, \delta \varepsilon_0 \cdot \eta_{p, 0} \right)
\,+\, \delta \varepsilon_0 \cdot \eta_{p, 0}
\end{array}
\right.
\label{TaE_1}
\end{eqnarray}

\noindent
and

\noindent
\begin{eqnarray}
\left\{
\begin{array}{ccc}
\delta x_{\varsigma} &=&\left(\delta x_{\varsigma}  \,-\, 
\delta \varepsilon_{\varsigma} \cdot \eta_{x, 0} \right)
\,+\, \delta \varepsilon_{\varsigma} \cdot \eta_{x, 0}\\
\delta p_{\varsigma} &=& \left(\delta p_{\varsigma}  \,-\, 
\delta \varepsilon_{\varsigma} \cdot \eta_{p, 0} \right)
\,+\, \delta \varepsilon_{\varsigma} \cdot \eta_{p, 0}
\end{array}
\right.
\label{TaE_2}
\end{eqnarray}

\noindent
where $\,\eta_{x, 0}\,$ and $\,\eta_{p, 0}\,$ are the coordinate and
momentum design dispersions respectively.

The first terms in the right hand sides of formulas (\ref{TaE_1}) and (\ref{TaE_2}) 
represent the pure betatron oscillations. Let us at the beginning estimate
their difference using the Courant-Snyder quadratic form, which in this case
will be an invariant, i.e. let us consider the random variable

\noindent
\begin{eqnarray}
\tilde{I}_x^{\varsigma} \;=\; I_x ( \,r, 
\,\left( \delta x_{\varsigma} -  \delta x_{0} \right)-
\left(\delta \varepsilon_{\varsigma} -  \delta \varepsilon_{0} \right)
\cdot \eta_{x, 0} \,,
\, \left(\delta p_{\varsigma}  -  \delta p_{0}\right) -
\left(\delta \varepsilon_{\varsigma} -  \delta \varepsilon_{0} \right)
\cdot \eta_{p, 0}\, ) . 
\label{TaE_3}
\end{eqnarray}

\noindent
The mean value of this variable is given below

\noindent
\begin{eqnarray}
\big< \, \tilde{I}_x^{\varsigma} \, \big> 
\;=\; 2 \,\epsilon_{\varsigma} \, m_p(\beta_{\varsigma},\,\beta_0) 
\;+\;\Delta_{\varsigma}^2 \cdot I_x\left(\,r,\,\eta_{x, \varsigma}-\eta_{x, 0}\,,\, 
\eta_{p, \varsigma}-\eta_{p, 0} \,\right)\,,
\label{TaE_4}
\end{eqnarray}

\noindent
and one sees that, in addition to the mismatch between error and design
betatron functions, the difference between error and design dispersions 
starts to play an important role.

If we again will assume that the random vector $\,\vec{\,\varsigma}\,$ 
has a multivariate normal distribution, we can represent
$\, \tilde{I}_x^{\varsigma} \,$ in the form 

\noindent
\begin{eqnarray}
\tilde{I}_x^{\varsigma} \;=\; \tilde{\mu}_{+} \, \tilde{\xi}_{+}^2 
\;+\; \tilde{\mu}_{-} \, \tilde{\xi}_{-}^2\,,
\label{TaE_5}
\end{eqnarray}

\noindent
which is similar to (\ref{SEC_GD_2}) and in which
$\,\tilde{\xi}_{\pm}\,$
are again two independent standard normal variables.
Unfortunately, the expressions for $\,\tilde{\mu}_{\pm}\,$
become essentially more complicated than (\ref{SEC_WC_11}) and,
with the notations

\noindent
\begin{eqnarray}
\hat{m}_p \,=\,m_p(\beta_{\varsigma},\,\beta_0),
\hspace{0.5cm}
\hat{I}_x\,=\,I_x\left(\,r,\,\eta_{x, \varsigma}-\eta_{x, 0}\,,\, 
\eta_{p, \varsigma}-\eta_{p, 0} \,\right),
\label{TaE_6}
\end{eqnarray}

\noindent
are given below

\noindent
\begin{eqnarray}
\tilde{\mu}_{\pm} =
\epsilon_{\varsigma} \,\hat{m}_p +
\frac{\Delta_{\varsigma}^2}{2} \cdot \hat{I}_x
\,\pm\,
\sqrt{\epsilon_{\varsigma}^2 \left(\hat{m}_p^2 \,-\, 1\right) + 
\epsilon_{\varsigma} \,\Delta_{\varsigma}^2 \left(\hat{m}_p - 1\right)\cdot\hat{I}_x
\,+\,\frac{\Delta_{\varsigma}^4}{4} \cdot \hat{I}_x^2\,}\,.
\label{TaE_7}
\end{eqnarray}

With representation (\ref{TaE_5}) one can calculate the variance

\noindent
\begin{eqnarray}
{\cal V} \big(\, \tilde{I}_x^{\varsigma} \, \big) \;=\;
\big< \, \big(\tilde{I}_x^{\varsigma}\big)^2 \, \big> 
\,-\, \big< \, \tilde{I}_x^{\varsigma} \, \big>^2 \;=\;
2\,\left(
\tilde{\mu}_{+}^2 \,+\,\tilde{\mu}_{-}^2
\right)
\label{TaE_8}
\end{eqnarray}

\noindent
and also find formula for the probability density $\,\tilde{p}(t)$.
This density is equal to zero for negative values of its argument,
and for $\,t \geq 0\,$

\noindent
\begin{eqnarray}
\tilde{p}(t) \;=\; \frac{1}{2 A} \,
I_0 \left(\sqrt{\,\left(\frac{B}{A}\right)^2 - 1\,} \,\cdot\, \frac{t}{2 A}\right)
\exp \left(-\frac{B}{A} \cdot \frac{t}{2 A}\right)\,,
\label{TaE_9}
\end{eqnarray}

\noindent
where $I_0$ is the modified Bessel function of zero order and

\noindent
\begin{eqnarray}
A \,=\, \sqrt{\,\tilde{\mu}_{+}\, \tilde{\mu}_{-}\,},
\hspace{1.0cm}
B \,=\, \frac{\tilde{\mu}_{+}\,+\, \tilde{\mu}_{-}}{2}
\label{TaE_10}
\end{eqnarray}

\noindent
are the geometric and the arithmetic means of the eigenvalues
(\ref{TaE_7}) respectively.

To finish this section, let us note that in order to get probabilistic
characteristic of the random variable (\ref{SEC_GD_1}), i.e. in order
to study not the difference in the pure betatron oscillations, but
the total difference in the transverse motion, one simply
has to set to zero the design dispersions in all formulas of
this subsection (independently, if actual design coordinate and
momentum dispersions are equal to zero or not).
The obtained formulas will, of course, not have invariant character
anymore. Nevertheless, the dependence form the position of the reconstruction
point will enter them through the single parameter, namely through the value
$\,I_x\left(\,r,\,\eta_{x, \varsigma}\,,\,\eta_{p, \varsigma}\,\right)$.


\begin{thebibliography}{99}

\bibitem{ThePaper} V.Balandin, W.Decking and N.Golubeva, 
``Errors in Reconstruction of Difference Orbit Parameters
due to  Finite BPM-Resolutions '', 
TESLA-FEL 2009-07, July 2009.

\bibitem{HamMethod} V.Balandin and N.Golubeva, 
``Hamiltonian Methods for the Study of Polarized
Proton Beam Dynamics in Accelerators and Storage Rings'', 
DESY 98-016, February 1998.

\bibitem{FLASH_1} V.Ayvazian et al.
``First operation of a free-electron laser generating GW power radiation at 32-nm wavelength'',
Eur.Phys.J.D37:297-303,2006.

\bibitem{FLASH_2} W.Ackermann et al.
``Operation of a free-electron laser from the extreme ultraviolet to the water window'',
Nature Photon.1:336-342,2007.

\bibitem{PedroBC2} P.Castro,
``Beam Trajectory Calculations in Bunch Compressors of TTF2'',
DESY Technical Note 03-01, April 2003.

\bibitem{GradSolomon} A.Grad and H.Solomon, 
``Distribution of quadratic forms and some applications'', 
Ann.Math.Statist., Vol.26 (1955).

\end{thebibliography}
\end{document}